  \documentclass{aa}  
\usepackage{graphicx}
\usepackage{natbib}
\usepackage{ulem}
\usepackage{txfonts}
\begin{document}
   \title{Internetwork magnetic field distribution from simultaneous $1.56$
$\mu$m and $630$ nm observations}

   \subtitle{}

   \author{M. J. Mart\' inez Gonz\'alez\inst{1}\thanks{\emph{Present address:}
LERMA, Observatoire de Paris-Meudon, 5 place de Jules Janssen, 92195, Meudon,
France}
          \and 
      M. Collados\inst{1}
      \and
      B. Ruiz Cobo\inst{1}
      \and
      C. Beck\inst{1}}

   \offprints{M. J. Mart\' inez Gonz\'alez}

   \institute{Instituto de Astrof\' isica de Canarias, V\' ia L\'actea S/N,
31200, La Laguna, Spain\\
             \email{Marian.Martinez@obspm.fr, mcv@iac.es, brc@iac.es,
cbeck@iac.es}}

   \date{Received ; Accepted }

 
  \abstract
   {}
   {Study the contradictory magnetic field strength distributions retrieved 
   from independent analyses of spectropolarimetric observations 
   in the near-infrared ($1.56$ $\mu$m) and in the visible (630 nm) 
   at internetwork regions.}
   {In order to solve this apparent controversy, we present simultaneous and
co-spatial $1.56$ $\mu$m and 630 nm observations of an internetwork area.
The properties of the circular and linear polarization signals, as well as the 
Stokes V area and amplitude asymmetries, are discussed.
As a complement, inversion techniques are also used to infer the physical 
parameters of the solar atmosphere. As a first step, the infrared and 
visible observations are analysed separately to check their 
compatibility. Finally, the simultaneous inversion of the two data 
sets is performed.}
   {The magnetic flux densities retrieved from the individual analysis of the
infrared and visible data sets are strongly correlated. The polarity of
the Stokes V profiles is the same at co-spatial pixels in both wavelength ranges.
This indicates that both \hbox{$1.56$ $\mu$m} and 630 nm observations trace the same
magnetic structures on the solar surface. The simultaneous inversion of the two
pairs of lines reveals an internetwork full of sub-kG structures that fill only 
2\% of the resolution element. A correlation is found between the magnetic field
strength and the continuum intensity: equipartition fields ($B\sim 500$ G) tend
to be located in dark intergranular lanes, whereas weaker field structures
are found inside granules. The most probable unsigned magnetic flux
density is \hbox{10 Mx/cm$^2$}. The net magnetic flux density in the whole field
of view is nearly zero. This means that both polarities cancel out almost exactly 
in our observed internetwork area.}
{}


   \maketitle
%


\section{Introduction}

Solar magnetic fields as seen in magnetograms form a particular pattern outside
active regions, in the so-called quiet Sun. The largest polarimetric signals are
confined to the supergranular boundaries forming the photospheric network.
Inside these large-scale convective cells it is possible to detect magnetic fields 
with a more disperse character and with smaller 
polarimetric signals. These areas are called the internetwork. The magnetic structures of the
network can be modeled either with a flux tube model or with the MISMA
hypothesis \citep[MIcroStructured Magnetic Atmosphere,][]{jorge_96}. Both
models reproduce the observed Stokes profiles and agree on the existence of nearly
vertical magnetic fields of kG strength occupying only some 10-20 \% of the
resolution element. Observations at $1.56$ $\mu$m
\citep[][]{muglach_92, lin_95}, \hbox{525 nm} \citep[][]{stenflo_73} and 630~nm
\citep[][]{jorge_00} give consistent results. Unfortunately, this is
not the case for the internetwork, with recent studies based on high quality 
spectro-polarimetric data at $1.56$ $\mu$m and $630$ nm yielding contradictory results. 

The first observation of Stokes V spectra of an internetwork region was carried
out by \cite{keller_94}. These authors observed the pair of Fe\,{\sc i} lines at
\hbox{$525$ nm} and analysed four average Stokes V spectra using the line ratio
technique \citep{howard_71}. Their conclusion was that the magnetic field strength
was below \hbox{$1$ kG (500 G)} with a probability of $95$\% (68\%). Later, 
\cite{lin_95} measured directly the Zeeman
splitting of the Fe\,{\sc i} line pair at $1.56$ $\mu$m to infer the magnetic
field strength. He showed that the typical magnetic field strength in the internetwork
was well below kG. The retrieved kG structures were correlated with network
patches. A few years later, \cite{lin_99}, using the same pair of infrared lines,
deduced field strengths between $0.2$ and $1$ kG.

The success of the Advanced Stokes Polarimeter \citep{lites_93} filled the
literature with works concerning the internetwork magnetism using the Fe\,{\sc
i} pair of lines at 630 nm. Most of these studies obtained that the field strength 
distribution at the internetwork presents a peak at kG. These kG fields,
even if they occupy a small fraction of the Sun's surface (0.1-1\%), contained most of
the magnetic flux of the internetwork. Applying a Principal Component Analysis
(PCA) inversion, \cite{hector_02} also recovered magnetic
field strengths of the order of kG. The database for this PCA algorithm consisted of synthetic
profiles generated under the MISMA hypothesis. The most surprising fact is that,
analysing the same data in terms of a Milne-Eddington inversion, \cite{lites_02}
obtained field strengths clearly below 1 kG. Using the line ratio technique,
\cite{ita_jorge_03} confirmed the presence of ubiquitous kG fields in the
internetwork. They used speckle reconstructed data in order to achieve a spatial 
resolution of $0.5''$. 

\cite{khomenko_03} carried out the study of an internetwork region at $1.56$
$\mu$m, measuring the Zeeman splitting to deduce the magnetic field strength. 
These authors obtained that the internetwork magnetic field strength
was well below the kG regime, with a probability distribution function that
could be reproduced by a decreasing exponential with an e-folding factor of $250$ G. 

From all these works, one can summarize that studies using the 630 nm or 
the 1.5 $\mu$m lines have led to contradictory
results: the infrared lines indicate fields below 1 kG, while the visible lines 
indicate a dominance of kG fields. Two plausible explanations have been presented
to explain such a discrepancy. One is based
on the idea that visible and near-infrared observations contain incompatible 
information about the internetwork magnetism \citep{hector03, jorge_ita_03}. The other 
suggests that the line pair at $630$ nm is not reliable for retrieving the magnetic
field strength at the internetwork \citep{luis_03, marian_06, arturo_07}. 

\cite{hector03} suggested that, if kG and sub-kG fields co-exist in the same resolution
element, the visible lines would be sensitive to the kG contribution, whereas
the infrared lines would trace the sub-kG structures. The explanation comes from
the different sensitivity of the spectral lines to the magnetic field. According
to \cite{luis_03}, this may be the situation in the case of having kG fields 
occupying $\sim$ 20\% of the resolution element and sub-kG fields filling the rest. 
If the percentages change, the premise seems not valid anymore. However, 
\cite{hector_04} found observational evidence indicating the coexistence of sub-kG 
and kG fields in the internetwork. Their analysis was based on a three-component
inversion, two of the components being magnetic with fixed field strengths of
1700 and 500 G and the other one being field-free. The observations were
compatible with $\sim$ 20\% of the resolution element filled by the \hbox{1700
G} field and the rest by the $500$ G field. 

\cite{luis_03} performed numerical simulations of Stokes profiles at 630 nm and
at 1.56 $\mu$m, assuming a magnetic field filling 5\% of the resolution
element. For consistency with the results by \cite{khomenko_03}, they parameterized
the field strength distribution as a decreasing exponential
with a mean value of 250 G. After adding some noise (to a level of
10$^{-3}$ I$_\mathrm{c}$, with I$_\mathrm{c}$ the continuum intensity) 
to the synthetic profiles and inverting them, the resulting magnetic field strength
distribution retrieved from the synthetic visible data set was centered at kG.
On the contrary, the distribution inferred from the synthetic infrared spectra 
was close to the original one. They concluded that, for a sufficiently small noise
level, both infrared and visible observations should agree and retrieve the
initial distribution. However, and advancing some of the results that we will show 
in this paper, other effects apart from the noise level make the \hbox{630 nm} 
magnetometry unreliable. This follows from the results by \cite{marian_06} that, 
for the present observing conditions (with magnetic flux
densities of $\sim$ 10 Mx/cm$^2$, noise level of \hbox{$6.5 \times 10^{-5}$ I$_\mathrm{c}$}, 
spatial resolution from 0.5$''$ to 1$^{\prime\prime}$), the 630 nm pair of lines does not carry enough information to retrieve the magnetic field strength. 

\begin{figure}
\centering
\includegraphics[width=7cm]{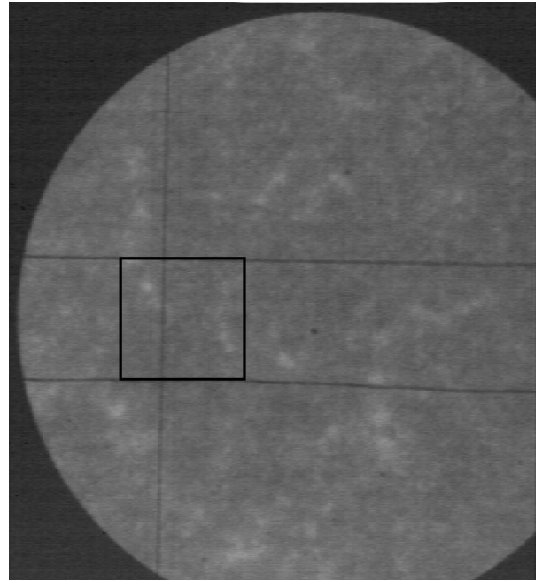}
\caption{Ca\,{\sc ii} K line core image from the VTT slit jaw system. Bright
features are good indicators of magnetic activity. The black square marks the
area scanned for this study. The lack of intense bright points in the observed
area indicates a very quiet region without intense magnetic activity.}
\label{slitjaw}
\end{figure}

Apart from numerical simulations, simultaneous and co-spatial observations in
both spectral ranges should give the final answer to the 1.56 $\mu$m-630 nm
controversy. The first work with this aim was carried out by \cite{jorge_ita_03} using
data from the Vacuum Tower Telescope (VTT) and THEMIS telescopes 
at the Observatorio del Teide. This
yielded two data sets in the $630$ nm and \hbox{$1.56$ $\mu$m} lines, but with a
significantly different spatial resolution since the THEMIS telescope was not
equipped with an image stabilization system at that time. The separate inversion
of both data sets with a Milne-Eddington approximation led to field strengths of
about $300$ G for the infrared lines and $1100$ G for the visible ones.
More surprisingly, $25$\% of the analysed signals showed opposite polarities 
in the same pixel at both wavelength ranges. \cite{khomenko_shelyag_05} showed, using 
magnetohydrodynamic simulations, that a different spatial resolution in both
wavelengths can result in apparent opposite polarities at the same spatial point. 
In a subsequent publication, \cite{ita_06} degraded the spatial
resolution of the VTT data used by \cite{jorge_ita_03} to match that of the
THEMIS observations.
From a simultaneous MISMA inversion they also obtained an
important fraction of kG fields in the visible and weaker fields in the
infrared. Consequently, they concluded that the information contained in the 
infrared data was incompatible to that contained in the visible data set. They assert 
that kG and weaker magnetic fields were coexisting in the resolution element, the visible 
being sensitive to the kG fields and the infrared to the weaker ones.

Recently, the polarimeter on-board HINODE has
provided spectro-polarimetric 
observations at 630 nm with the highest spatial resolution ever achieved with
this kind of data (0.3$''$). If 
the occupation fraction of the magnetic elements in the resolution element would have increased at least one order of magnitude from 1$''$ to 0.3$''$ one would have been capable of separating the effects of the thermodynamics and the magnetic field strength \citep{andres_marian_07_astroph}. 
\cite{david_07_astroph} perform a Milne-Eddington inversion of HINODE's data on a quiet Sun region. The noise level is around one order of magnitude higher than the one of the observations presented in this paper. In order to perform the inversions they assume one single magnetic atmosphere in the resolution element and some contamination of stray light. The retrieved magnetic field strength distribution has a peak at very weak fields, in disagreement with the 
previous results using the pair of spectral lines at 630 nm. However, even if this result is compatible with the magnetic fields obtained using the infrared lines at 1.5 $\mu$m, the validity 
of the visible spectral lines at the HINODE observational conditions should be tested.

In this work we present simultaneous and
co-spatial $1.56$ $\mu$m and 630 nm observations taken {\it at the same telescope,
under very similar seeing conditions and using image stabilization systems}. 
Spurious effects arising from a different spatial resolution in both data sets, 
due to different observing conditions, are thus avoided. Prior to the analysis, 
noise is reduced to very low values using a PCA procedure to ensure that it does 
not bias the study of the compatibility of the internetwork observations at these two
wavelengths.



\begin{figure}
\includegraphics[width=4.3cm]{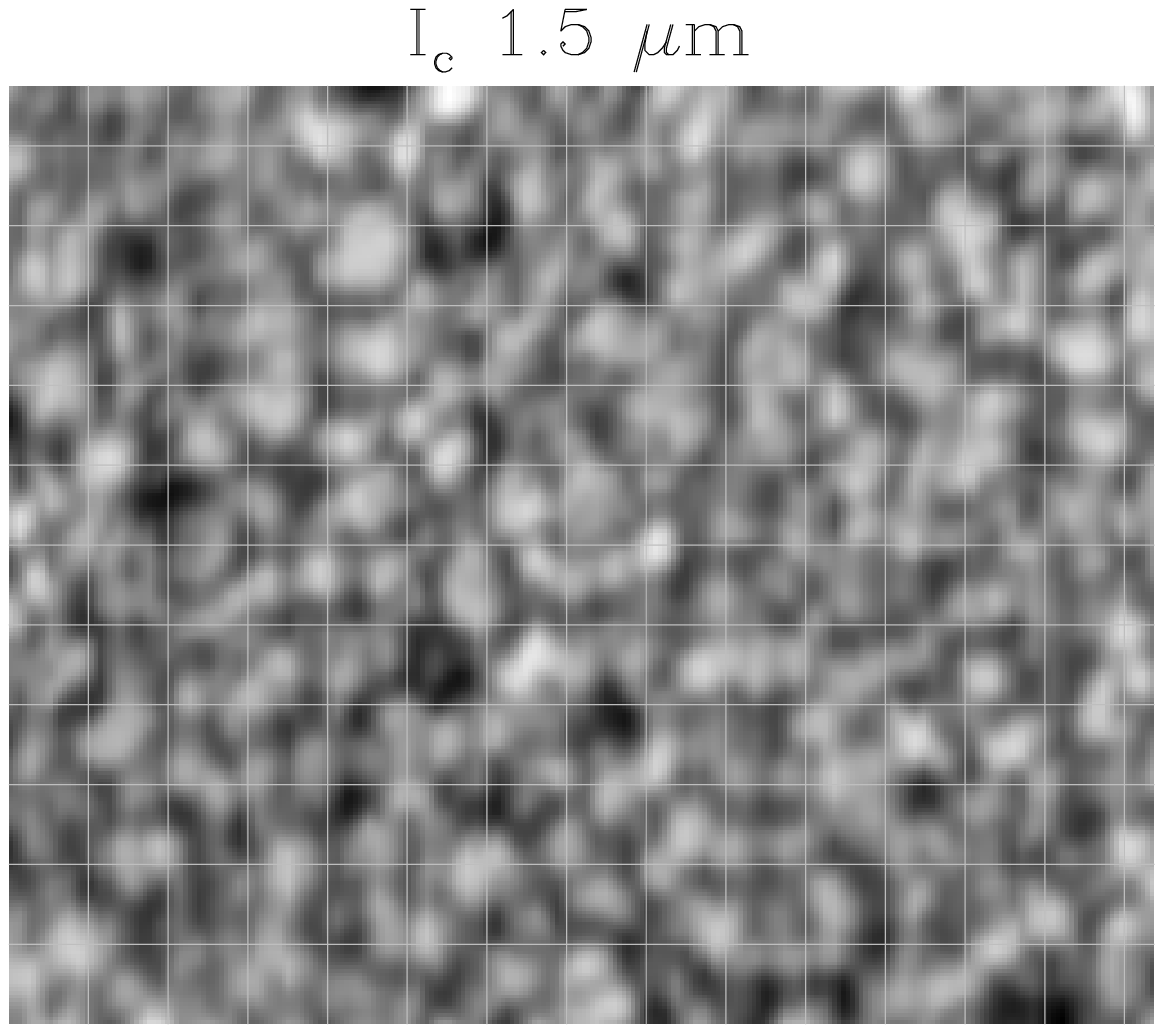}
\hspace{0.3cm}
\includegraphics[width=4.3cm]{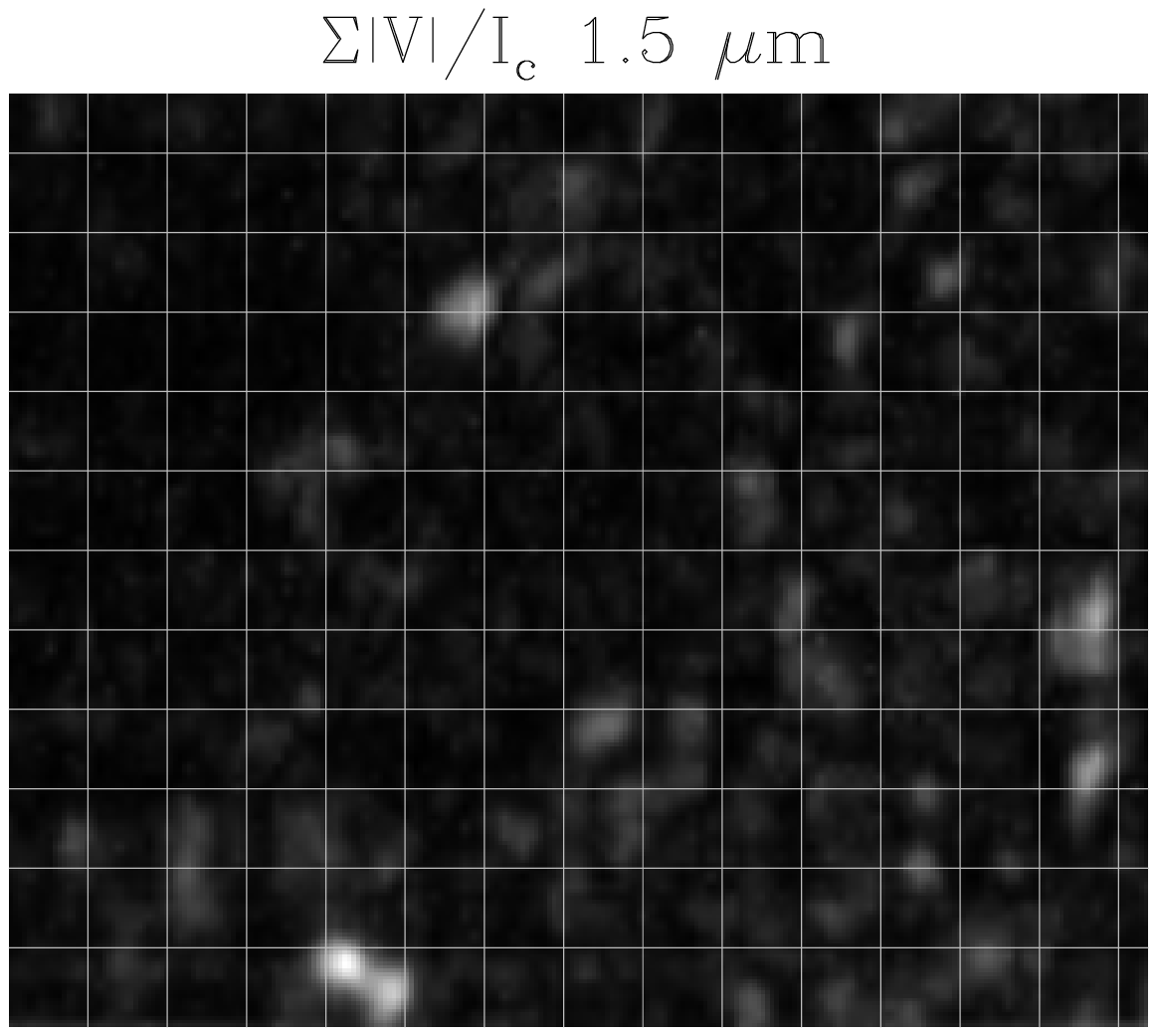}\\
\\
\includegraphics[width=4.3cm]{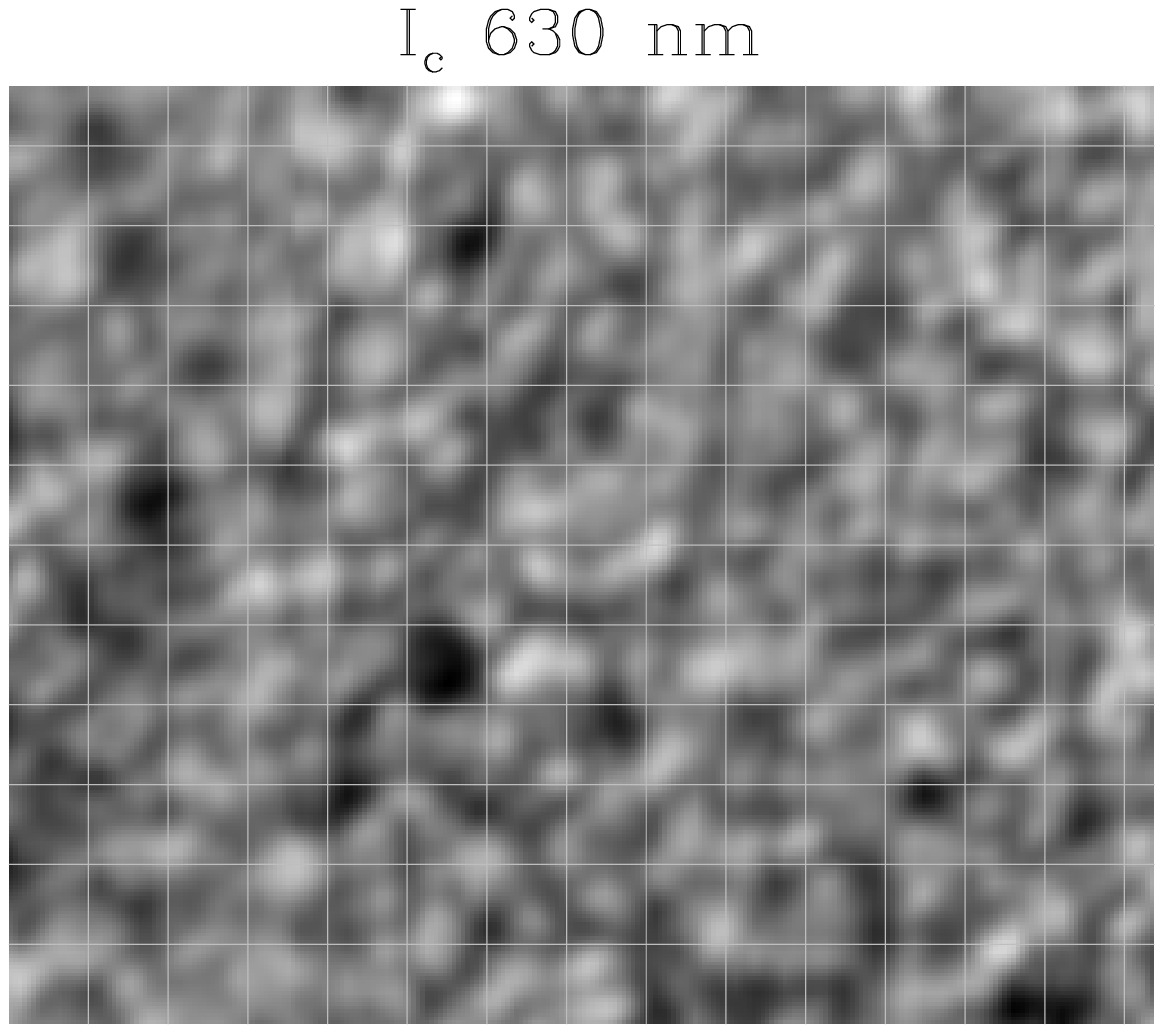}
\hspace{0.3cm}
\includegraphics[width=4.3cm]{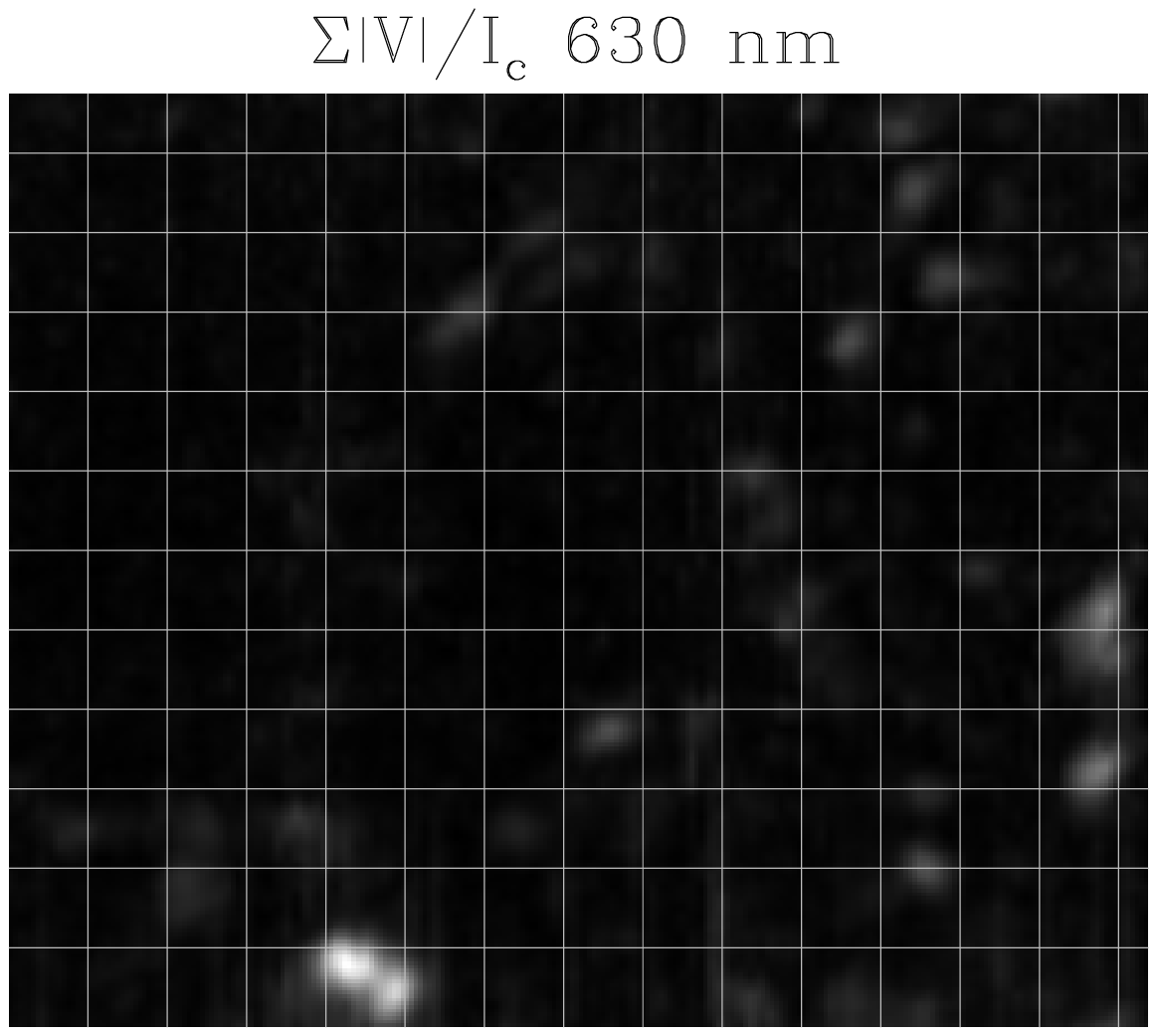}
\caption{Maps of continuum intensity and integrated absolute circular 
polarization for the aligned infrared (top) and visible (bottom) data. 
A grid is superposed for an easier visibility of the co-spatiality of structures.}
\label{mapas_alineados}
\end{figure}

\section{Observations}

On August 17, 2003, a very quiet
region at disc centre was observed at the VTT of the Observatorio 
del Teide from UT 07:34:14 to 08:26:31. A two-dimensional map was obtained 
by stepping the solar image perpendicularly to the spectrograph
slit. At each point, the full Stokes vector was recorded at the \hbox{$1.56$ $\mu$m} and
\hbox{630 nm} spectral ranges simultaneously. The light coming from the telescope was
divided using an achromatic 50\% - 50\% beam splitter, which sent half of the
incoming radiation to the infrared polarimeter TIP \citep{manolo99} and the
other half to the visible polarimeter POLIS \citep{beck_05}. 

In order to study the internetwork, network regions or other
activity areas on the solar surface were explicitly avoided with the aid of live
Ca\,{\sc ii} K images, which were accessible during the observations. 
The emission in the core of this spectral line is a good indicator of the magnetic
activity \citep[e.g.,][]{lites_99}. In Fig.~\ref{slitjaw} the observed region as
seen in the Ca\,{\sc ii} K line is shown. The black square contains the scanned area and
the vertical line inside it represents the position of the slit at a given time.
There were no intense bright zones in the whole scanned region, indicating that
there was no significant magnetic activity. The size of the scanned area was
33.25$''$ (along the slit) $\times$ 42$^{\prime\prime}$ (along the scan
direction). The image was stabilized using a correlation tracker system
\citep{ballesteros_96}, which allowed accurate stepping by 0.35$''$. The
sampling of the spectral images was 0.35$''$ $\times$ 29.6 m\AA\ 
at 1.56 $\mu$m and 0.14$''$ $\times$ 14.9 m\AA\ at 630~nm. The integration time
at each slit position was around \hbox{27 s}. The slit widths corresponded to 0.35$''$
for TIP and 0.48$''$ for POLIS. The spatial resolution was estimated
to be about 1.2$''$ and 1.3$''$ in the infrared and visible spectral ranges,
respectively. The continuum contrast of the granulation were $1.3$\% (infrared)
and $2.7$\% (visible).


\section{Data reduction}

Dark counts subtraction, flat-field correction and polarimetric
demodulation were performed with software packages available
for both instruments. Most of the instrumental polarisation crosstalk was 
removed using calibration optics located before the beam splitter
\citep{schlichenmaier_02}. For a complete correction, the coelostat 
configuration of the telescope needs to be modeled \citep{manolo99, 
beck_schlichenmaier_05}. The residual
crosstalk from Stokes I to Stokes Q, U and V was removed as:
\begin{equation}
\bf{P'}=\bf{P}-\gamma I ,
\end{equation}
being $\bf{P}$ or $\bf{P'}$ either Stokes Q, U or V. The symbol $I$ accounts for the intensity 
profile. The $\gamma$ coefficient is obtained by forcing the
continuum of the polarization profiles to zero since we do not expect continuum 
polarization due to the Zeeman effect. Then, $\gamma$ can be computed as:
\begin{equation}
\gamma = \frac{P_c}{I_c},
\end{equation}
being $P_c$ and $I_c$ the values the continuum of the Stokes Q, U or V profiles and of the intensity profiles, respectively. The residual crosstalk between
Stokes Q, U and V is very difficult to remove. Mainly in the 630 nm data, a few
percent may still remain. 

After this standard data reduction, two key items still remained to correct 
for a precise and reliable analysis: the differential
refraction of the Earth's atmosphere and the noise level.

\paragraph{Differential refraction and data alignment} The differential
refraction of the Earth's atmosphere leads to a time-dependent spatial
displacement of the solar images in different wavelengths, whose amount and
direction depend on the combined geometrical/optical configuration of the telescope and
instruments \citep{reardon_06}. We already tried to minimise refraction
effects during the observations by manually moving the POLIS scan mirror
correspondingly in-between exposures. In order to correct for the residual differential
diffraction, the continuum images of both spectral ranges were aligned. 
To that aim, the continuum intensity map from TIP data was divided into 24
small slices with a size of $33.25''\times1.75''$. The POLIS continuum image was divided 
into slices of $45.64''\times1.75''$. The correlation
coefficient $r(d_x,d_y)$, for integer pixel displacements, was calculated as
\begin{equation}
r(d_x,d_y)=\frac{\sum_{ij}[A_T(i+d_x,j+d_y)-\overline{A_T}][A_P(i,j)-\overline{
A_P}]}{\sqrt{\sum_{ij}[A_T(i+d_x,j+d_y)-\overline{A_T}]^2}\sqrt{\sum_{ij}[A_P(i,
j)-\overline{A_P}]^2}},
\end{equation}
where $\overline{A_T}$ and  $\overline{A_P}$ are the average intensities of the matching areas 
between the TIP and POLIS slices, respectively, and $d_y$ and $d_x$ denote displacements along 
and perpendicular to the
slit, respectively. The position of maximum correlation coefficient gives the
required displacement to have co-spatial granulation maps. These displacements 
inferred from the continuum
images were then used to retrieve the co-spatial spectra at each pixel. The
displacement perpendicular to the slit was always smaller than $0.6''$.
Figure \ref{mapas_alineados} shows the aligned maps. It is clear that each
recognisable feature is located at the same position in both the $1.56$ $\mu$m
and the $630$ nm continuum intensity (or Stokes V) map.

\paragraph{Noise level} The noise level reached during the observations was 
$2\times 10^{-4}$ and $2.5\times 10^{-4}$ I$_\mathrm{c}$ for the $1.56$ $\mu$m and 
630 nm data, respectively, with 
I$_\mathrm{c}$ being the mean continuum intensity. This noise level is typical for
spectro-polarimetric data. The expected signals for internetwork fields are however so weak
($\ll$ $10^{-3}$ I$_\mathrm{c}$) that we tried to improve as much as possible
the signal-to-noise ratio of our data. In order to reduce the noise level, we used a
statistical de-noising procedure based on the Principal Component Analysis (PCA) \citep{rees03}.
The main idea is to separate the noise contribution (or residual
features from the reduction process) from the real polarimetric information
as follows. 

Each one of our $n$ observed profiles is assumed to be a vector of size $p$, 
the number of wavelength points. In principle, each observed profile
would be represented by a point in a $p$-dimensional space, so that
one would need $p$ parameters to fully describe it.
If correlations exist between these
parameters, the cloud that represents all the observed profiles in the $p$-dimensional space
must be elongated in some particular directions. If the matrix $\hat{O}$
contains the $n\times p$ observed profiles, these directions
of maximum correlation can be obtained by diagonalising the correlation matrix of the
observations \citep{casini03}, defined as follows:
\begin{equation}
\hat{X}=\hat{O}^t\cdot \hat{O},
\end{equation}
where $\hat{X}$ is the $p\times p$ correlation matrix, the operator $\cdot$
indicates the matrix product and $t$ denotes matrix transposition. The
eigenvectors of the correlation matrix form a suitable basis in which the
observations can be decomposed as follows:
\begin{equation}
\hat{C}=\hat{O}\cdot \hat{B},
\end{equation}
being $\hat{C}$ the $n\times p$ coefficient matrix and $\hat{B}$ the $p\times p$ 
matrix containing the eigenvectors of the correlation matrix. Each $C_{ij}$ is the 
projection of the $i$-th observation on the $j$-th eigenvector.

The election of this basis (the principal components) is not casual. The
correlation matrix is constructed from solar profiles and the eigenvectors may
have a physical meaning \citep{skumanich_02}. However, the most important property
of the PCA decomposition for our purposes
is that, sorting the eigenvalues in descending order by their value, their
magnitude rapidly decreases. This means that the first few
eigenvectors contain the majority of the information. 
Under the assumption that the relevant information is
contained in the eigenvectors with higher eigenvalues, the PCA basis can be truncated
and those eigenvectors with small eigenvalues can be eliminated. With this, one ends up 
with a set of $p'$ eigenvectors that
contain mainly useful solar information. The data can be reconstructed as
\begin{equation}
\hat{O}_\mathrm{filt}=\hat{C'} \cdot \hat{B'}^{-1} ,
\end{equation}
where $\hat{O}_\mathrm{filt}$ is the $n\times p$ matrix with the filtered data set, $\hat{C'}$ is
the $n\times p'$ coefficient matrix and $\hat{B'}$ is the $p\times p'$ pseudo-basis 
matrix that contains only the $p'$ first
eigenvectors. The de-noising procedure is very easy, but the election of the new
basis to represent the data set is a tricky point. In an ideal case, one is dealing with 
spectro-polarimetric information contaminated by an additive noise.
In this situation, the PCA decomposition separates perfectly the noise and the real
information, giving the smaller eigenvalues to the eigenvectors dominated by
noise. In a real case, there exist unfortunately some other sources of noise
(for instance, interference fringes, cross-talk from one polarisation
state to the others, bad pixels, residual features of telluric lines on the
polarization profiles, etc). All these contributions to the observed spectra have
some kind of pattern and are not compatible with uncorrelated noise. The fact 
that they present clear patterns means that sometimes those spurious signals 
appear mixed with spectro-polarimetric 
information with non-negligible eigenvalues. Thus, the election
of the new basis partly has a subjective character. We kept only the first 16 
eigenvectors for the \hbox{1.5 $\mu$m} data and the 31 first ones for the 630 nm data. 
In both cases the eigenvalues of the rejected profiles are more than 3 orders of 
magnitude smaller than the maximum eigenvalue. Recently, \cite{andres_hector_07} have 
proposed a numerical procedure to find the optimum number of eigenvectors for
such kind of problem. Our selection of the reduced basis is in agreement with
their calculations.

\begin{figure}
\centering
\includegraphics[width=9cm]{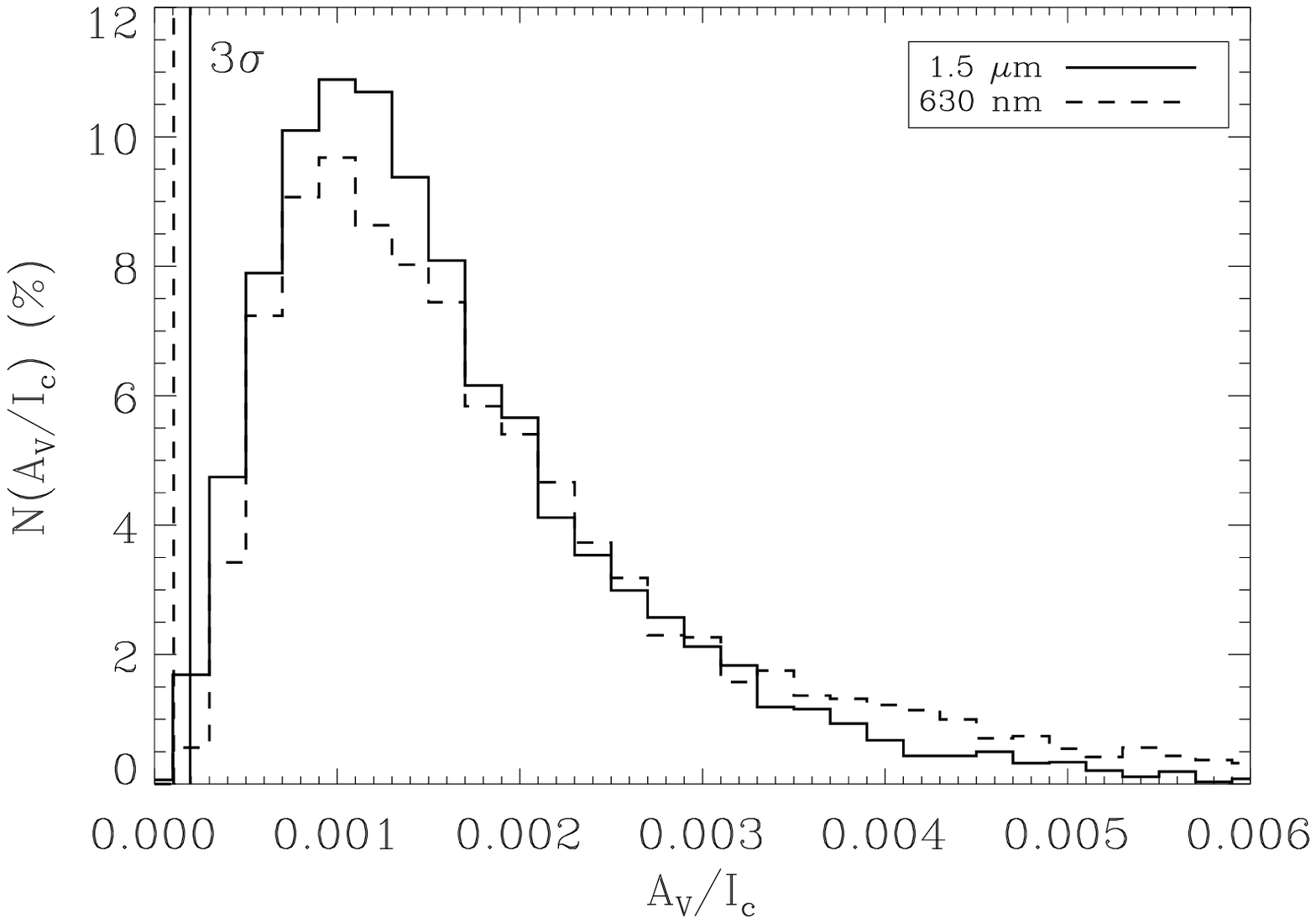}\\
\includegraphics[width=9cm]{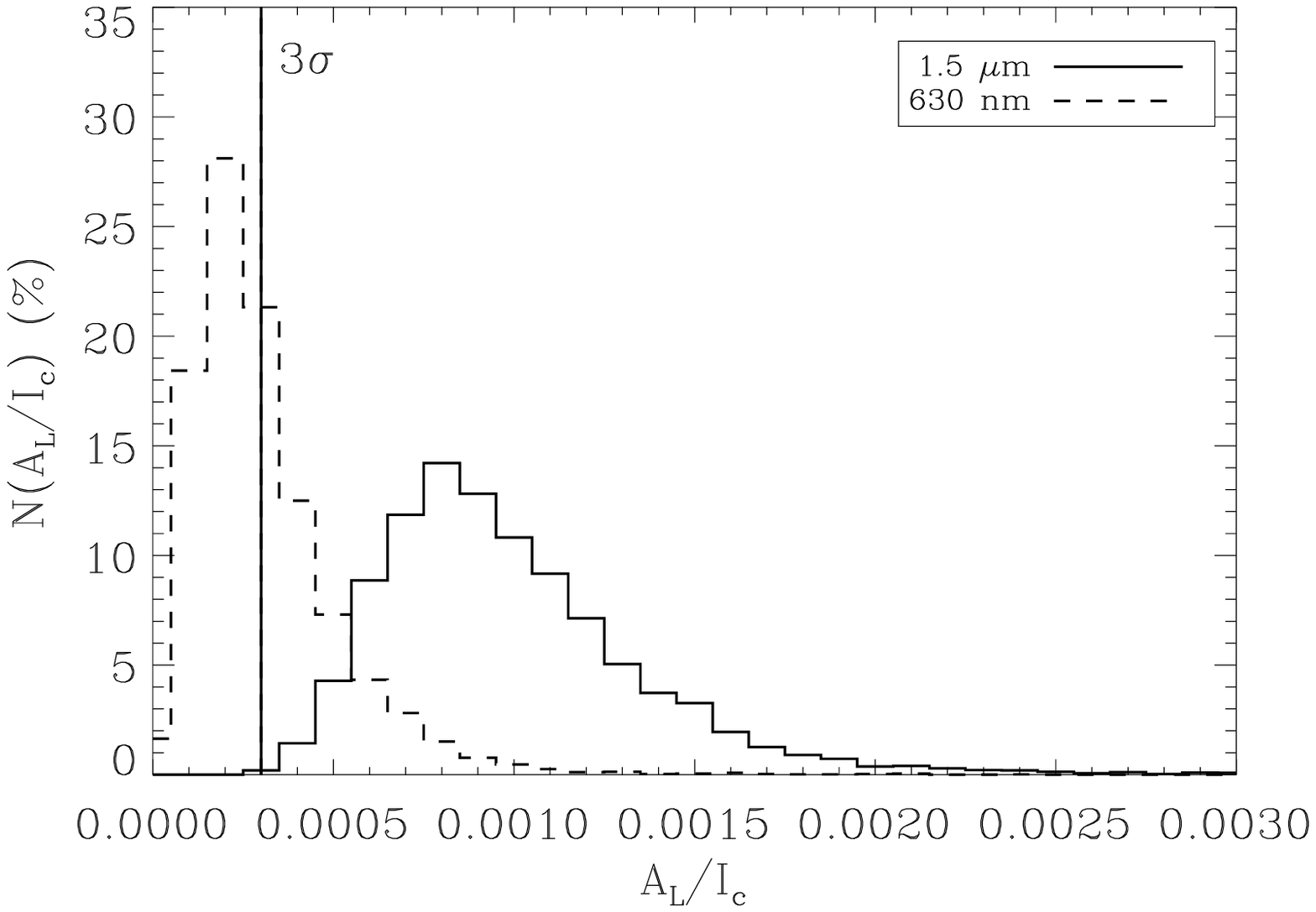}\\
\includegraphics[width=9cm]{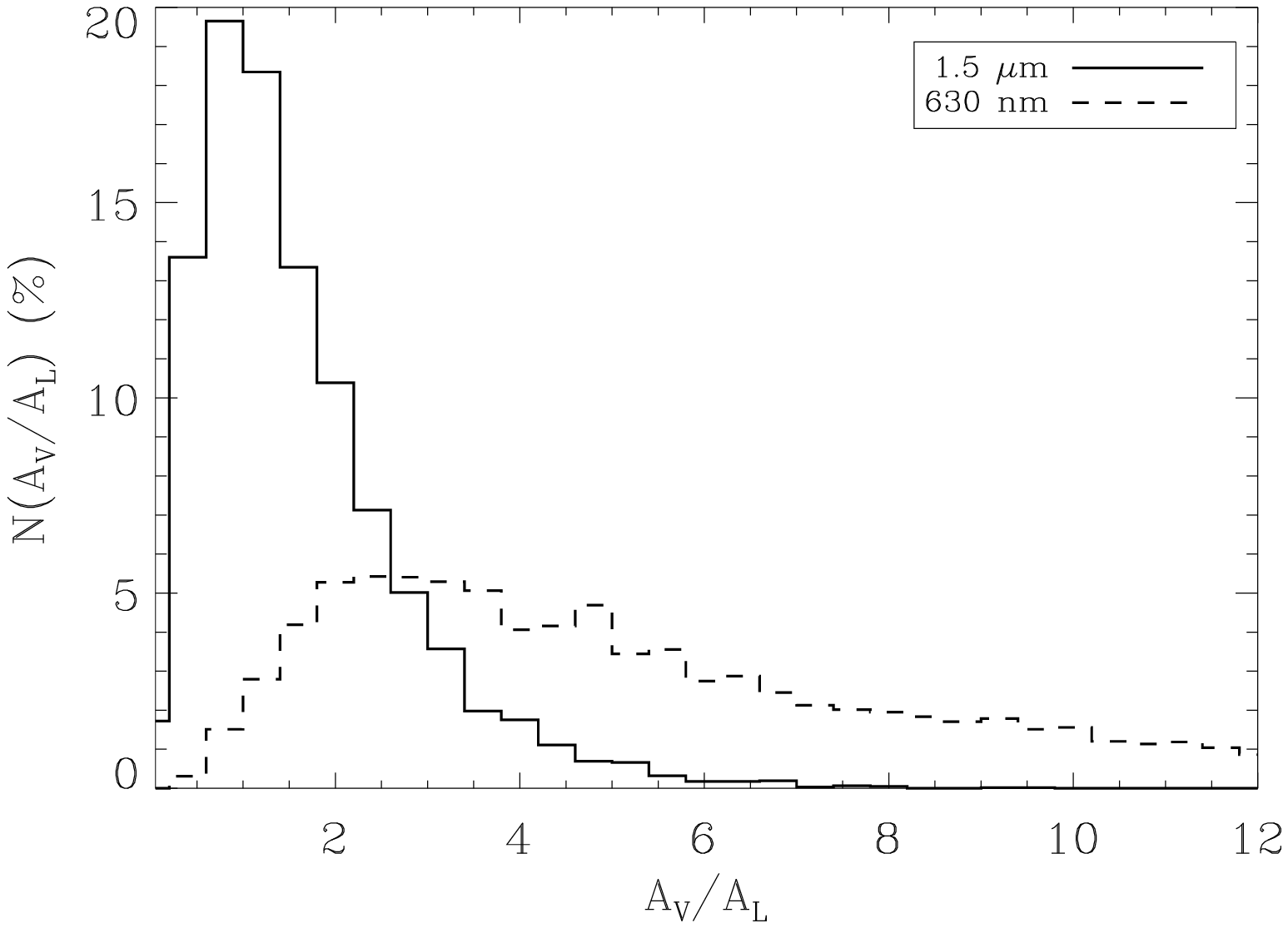}
\caption{Top panel: circular polarization amplitude of the 1564.8~nm line (solid line)
and the 630.25~nm one (dashed line). Middle panel: linear polarization amplitude for the
same spectral lines. Bottom panel: ratio of the circular and linear polarization
amplitudes for the same spectral lines.}
\label{circ_lineal_ratio}
\end{figure}

\begin{figure*}
\centerline{
\includegraphics[width=9cm]{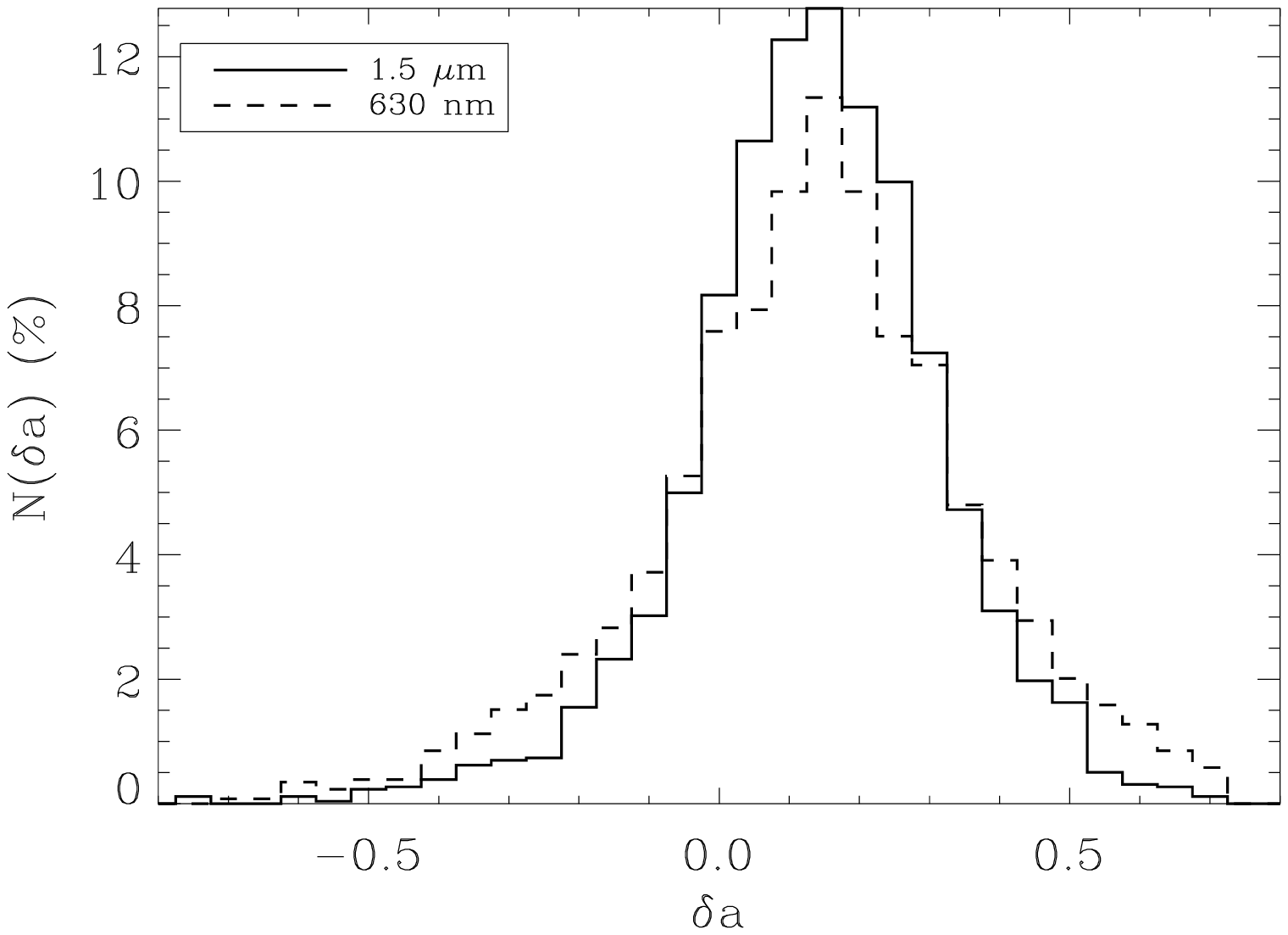}
\includegraphics[width=9cm]{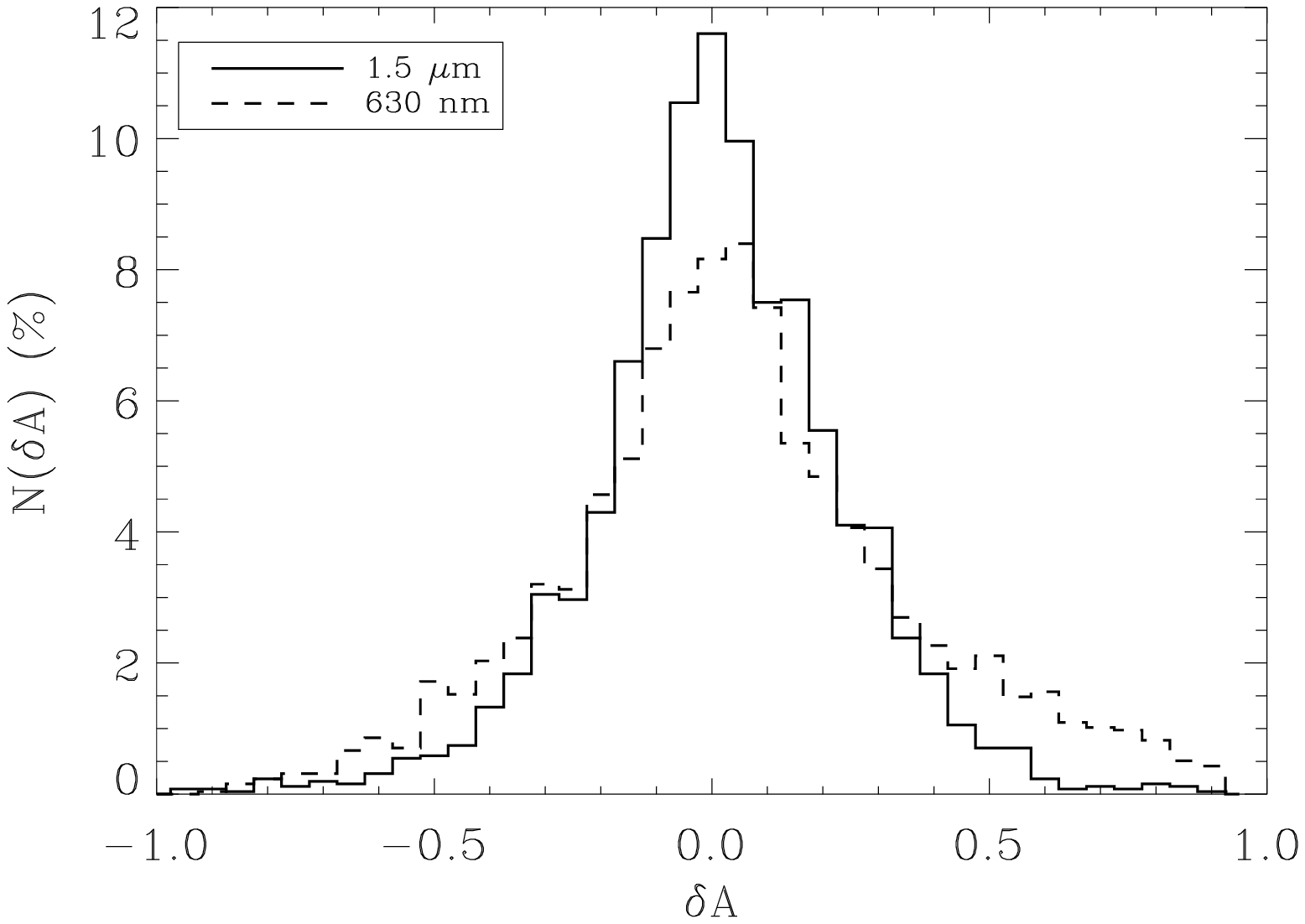}}
\caption{Histograms of the amplitude (left) and area (right) asymmetries of the
two-lobed Stokes V profiles at $1.5648$ $\mu$m (solid line) and at \hbox{630.2 nm} (dashed
line).}
\label{asim_tip_polis}
\end{figure*}

After applying this procedure, a filtered set of spectra with a very low noise level
is obtained. The Stokes V profiles show noise values of $4\times10^{-5}$ and $7\times 10^{-5}$
I$_\mathrm{c}$ at $1.56$ $\mu$m and 630 nm, respectively. The noise in the linear
polarization spectra is \hbox{$10^{-4}$ I$_\mathrm{c}$} in both data sets. 
%


\section{Analysis of polarization profiles}

\subsection{Amplitude of circular and linear polarization}

In this section we focus on the study of directly measurable polarimetric
quantities: the amplitudes of circular and linear polarization as well as the
amplitude and area asymmetries of the Stokes V profiles. Interestingly, 92.6\% of the observed
area showed signals above three times the noise level, indicating that the whole
surface is full of magnetic fields. However, in order to make a reliable analysis we
selected those profiles that fulfill, simultaneously in
both spectral ranges, the requirement that the maximum total polarization degree is larger than
$10\sigma$, with $\sigma$ the standard deviation of the noise level at continuum
wavelengths. This translates into $4\times 10^{-4}$ for the infrared and $7\times 10^{-4}$ I$_\mathrm{c}$
for the visible. In order not to bias our results, the amplitude of total polarization
$A_p$ was computed for the line with less magnetic sensitivity in each spectral range (1.5652 $\mu$m and
\hbox{630.15~nm}) as the maximum of $\sqrt{Q^2+U^2+V^2}$. This gives 56.4\% of the field of view
with total polarization signal over the $10\sigma$ threshold.

Although the profile selection was carried out with the spectral lines with the
weakest signals, we computed the amplitudes of 
circular and linear polarization with the lines with the largest magnetic sensitivity to the
magnetic field (1.5648 $\mu$m and \hbox{630.25 nm}). 
The amplitude of circular polarization ($A_V$) has been calculated as the maximum of
Stokes V and the amplitude of linear polarization ($A_L$) as the maximum of $\sqrt{Q^2+U^2}$. Figure
\ref{circ_lineal_ratio} shows the resulting histograms. The circular polarization distribution is
very similar in both spectral ranges, although small differences exist. The visible
lines detect slightly less weak signals than the infrared ones and slightly
larger amplitudes. There is also an interesting feature common to both spectral
ranges: the peak of the distributions is clearly above the noise level. One would
expect that undetected weak signals (due to the noise level or to the sensitivity 
of the detectors) should lead to maxima of the distributions at noise level.
\cite{wang_95} studied the internetwork
magnetism at a spatial resolution of $\sim$ 2$''$ using Big Bear deep
magnetograms. They observed a magnetic flux distribution that 
peaked at $5\times 10^{16}$ Mx \hbox{($\sim$ 10 Mx/cm$^2$)}, far from their
detection threshold of $10^{16}$ Mx \hbox{($\sim$ 2 Mx/cm$^2$)}. They examined seeing
and selection effects on their results and concluded that they were not
responsible for the peak and the consequent drop towards the smallest values of
the distribution down to the detection limit. We have also performed numerical
simulations to study the solar nature of this peak above the detection limit. In our case, we
obtain that cancellations due to the poor spatial resolution produce a similar
drop towards the detection limit. This means that the observed peak well above
the noise level might be an observational evidence of cancellations in the
resolution element.

The linear polarization histograms show very different distributions in the
two spectral ranges. The visible lacks many of the signals that are detected in the infrared. 
Of importance is to note that the Stokes Q and U signals observed in the infrared present
amplitudes that are of the same order of magnitude as those found in Stokes V. The peak of the
visible distribution of $A_L$ is close to the noise level of $10^{-4}$ I$_\mathrm{c}$, and
almost no signal is detected for values of $A_L$ above $10^{-3}$ I$_\mathrm{c}$, just the value for
which the infrared distribution reaches its maximum. The infrared distribution clearly presents
a peak for values much higher than the $3\sigma$ level, as already found for the amplitude
of circular polarization.

In order to shed some light into the previous results, it is interesting to recall the quantities
introduced by \cite{egidio} that give an idea of the sensitivity of
a spectral line to the circular and linear polarization. These quantities are defined as:
\begin{eqnarray}
s_V&=&\frac{\lambda}{\lambda_{ref}}\overline{g}d_c  \nonumber \\
s_L&=& \left( \frac{\lambda}{\lambda_{ref}} \right) ^2 \overline{G} d_c,
\end{eqnarray}
where $s_V$ and $s_L$ are the circular and linear polarization sensitivity index, respectively. 
The symbol $\lambda_{ref}$ stands for a reference wavelength.
Consequently, these numbers have only sense when comparing two different spectral
lines. The symbol $\overline{g}$ stands for the effective Land\'e factor of the
transition and $d_c$ is the central depression in terms of the continuum
intensity:
\begin{equation}
d_c=\frac{I_c-I(\lambda_0)}{I_c},
\end{equation}
with $I(\lambda_0)$ being the intensity at the minimun of the line minimum. The symbol
$\overline{G}$ plays the same role as the effective Land\'e factor but for linear
polarization. It is defined as:
\begin{equation}
\overline{G}=\overline{g}^2-\delta,
\end{equation}
where $\delta$ is a quantity that depends on the quantum numbers of the
transition \cite[see][]{egidio}. In both the 1564.8~nm and 630.25~nm spectral
lines we have $\delta=0$. The ratios between the circular and linear
polarization indices of the infrared and visible spectral lines are:
\begin{eqnarray}
\frac{(s_V)_{1.5 \mu \mathrm{m}}}{(s_V)_{630 \mathrm{nm}}}&=&1.37\nonumber \\
\frac{(s_L)_{1.5 \mu \mathrm{m}}}{(s_L)_{630 \mathrm{nm}}}&=&4.17.
\end{eqnarray}
This means that the sensitivity to the circular polarization is similar in both
wavelengths but the sensitivity to the linear polarization is 4 times larger at 1.5 $\mu$m than at 630 nm.
This result can explain the observed behaviour: the presence of ubiquitous
linear polarization signals in the infrared and the lack of them in the visible.

The bottom panel of Fig.~\ref{circ_lineal_ratio} shows the ratio of the circular
and linear polarization amplitudes. The histogram of the \hbox{630.25 nm} line shows a
broad maximum near $\sim 3$, i.e.~the majority of the circular polarization
signals are 3 times larger than the linear polarization signals. The 1.5 $\mu$m
histogram has a peak at $\sim 1$, i.e.~the linear polarization signals are as
large as the circular ones. If we compute the theoretical ratio, $s_V/s_L$, of
the infrared and the visible lines, we obtain:
\begin{equation}
\frac{(s_V/s_L)_{1.5 \mu \mathrm{m}}}{(s_V/s_L)_{630 \mathrm{nm}}}=0.34.
\end{equation}
Our observational data show a ratio between the circular and linear
polarization of the infrared relative to the visible of 0.33, close to the
theoretical prediction. As the formulae used are based 
on the weak field approximation, this indicates that the magnetic fields that
give rise to the observed 
signals are most probably in the weak field regime. This means that the majority
of the field strengths may be below $\approx 600$ G. 

\begin{figure*}[!t]
\centering
\includegraphics[width=8cm]{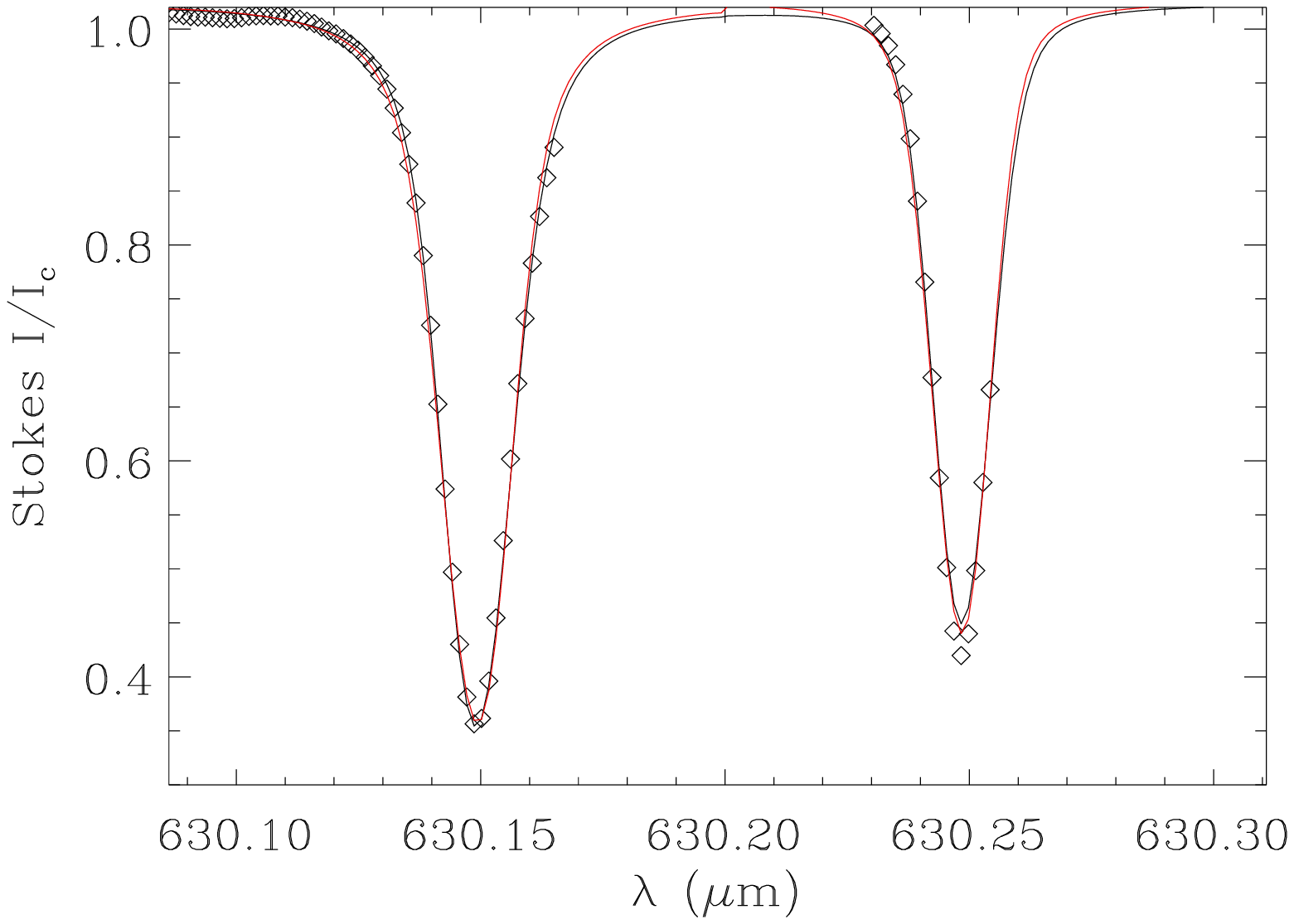}%
\includegraphics[width=8cm]{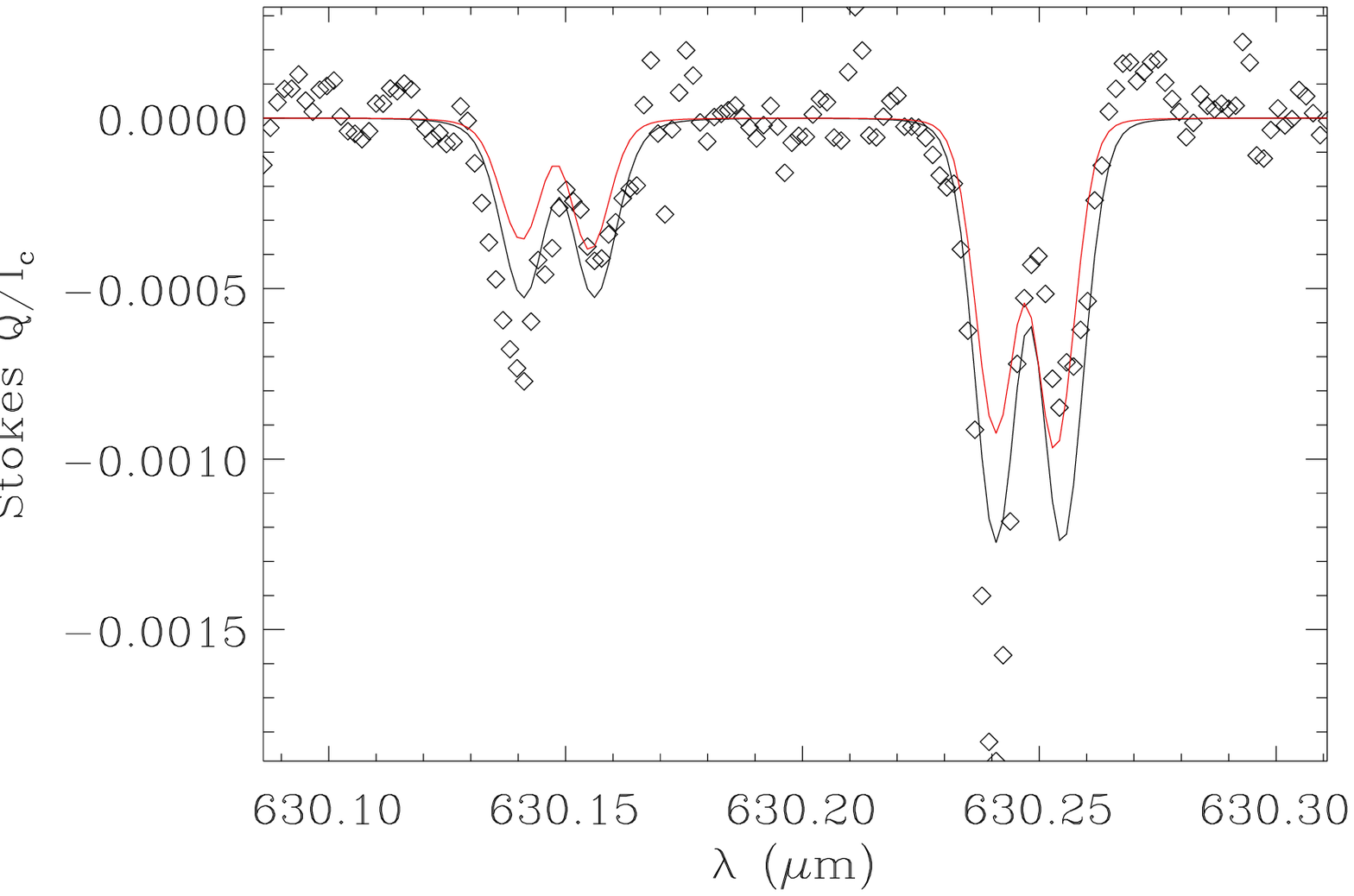}\\
\includegraphics[width=8cm]{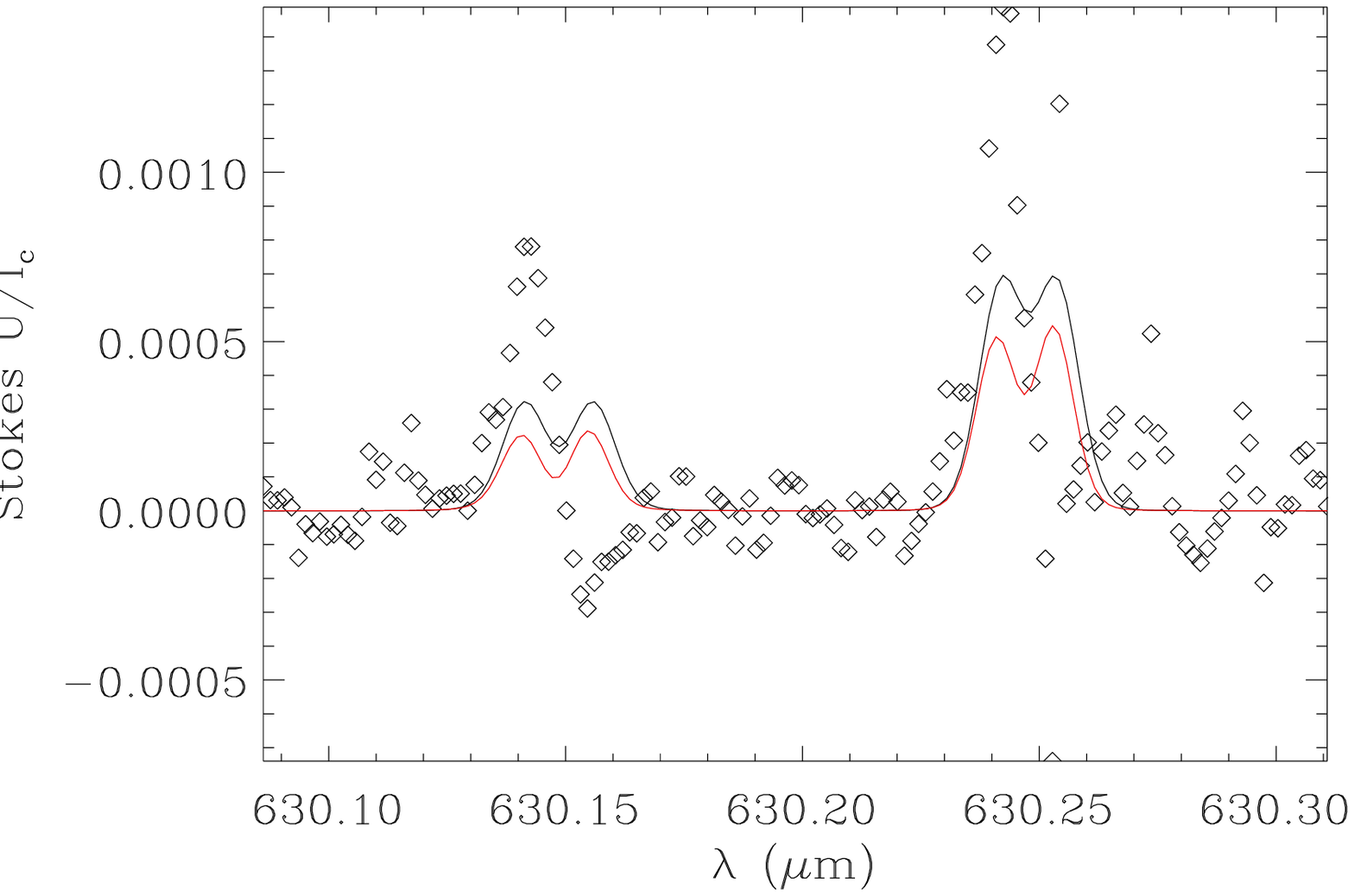}%
\includegraphics[width=8cm]{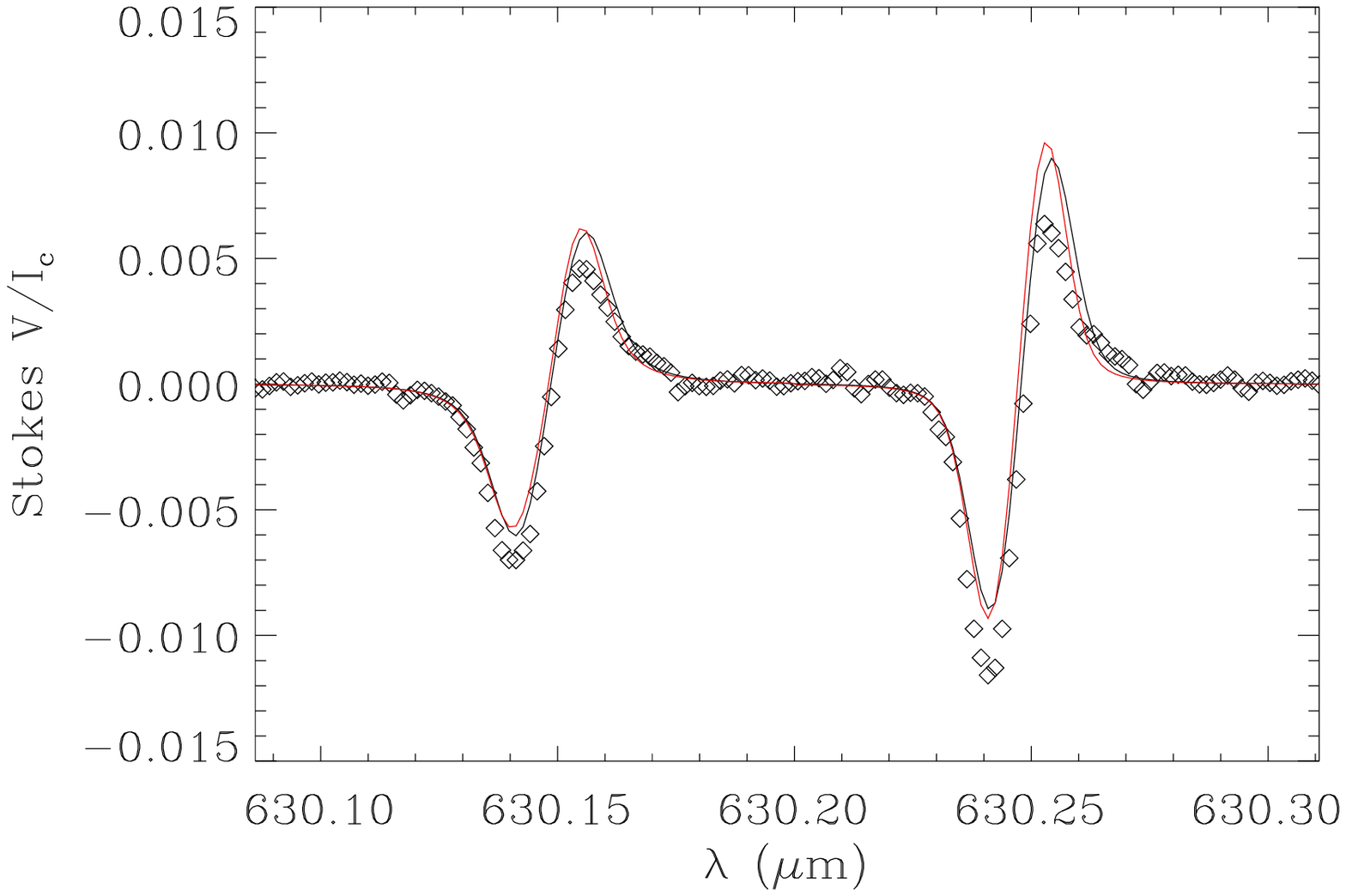}\\
\includegraphics[width=8cm]{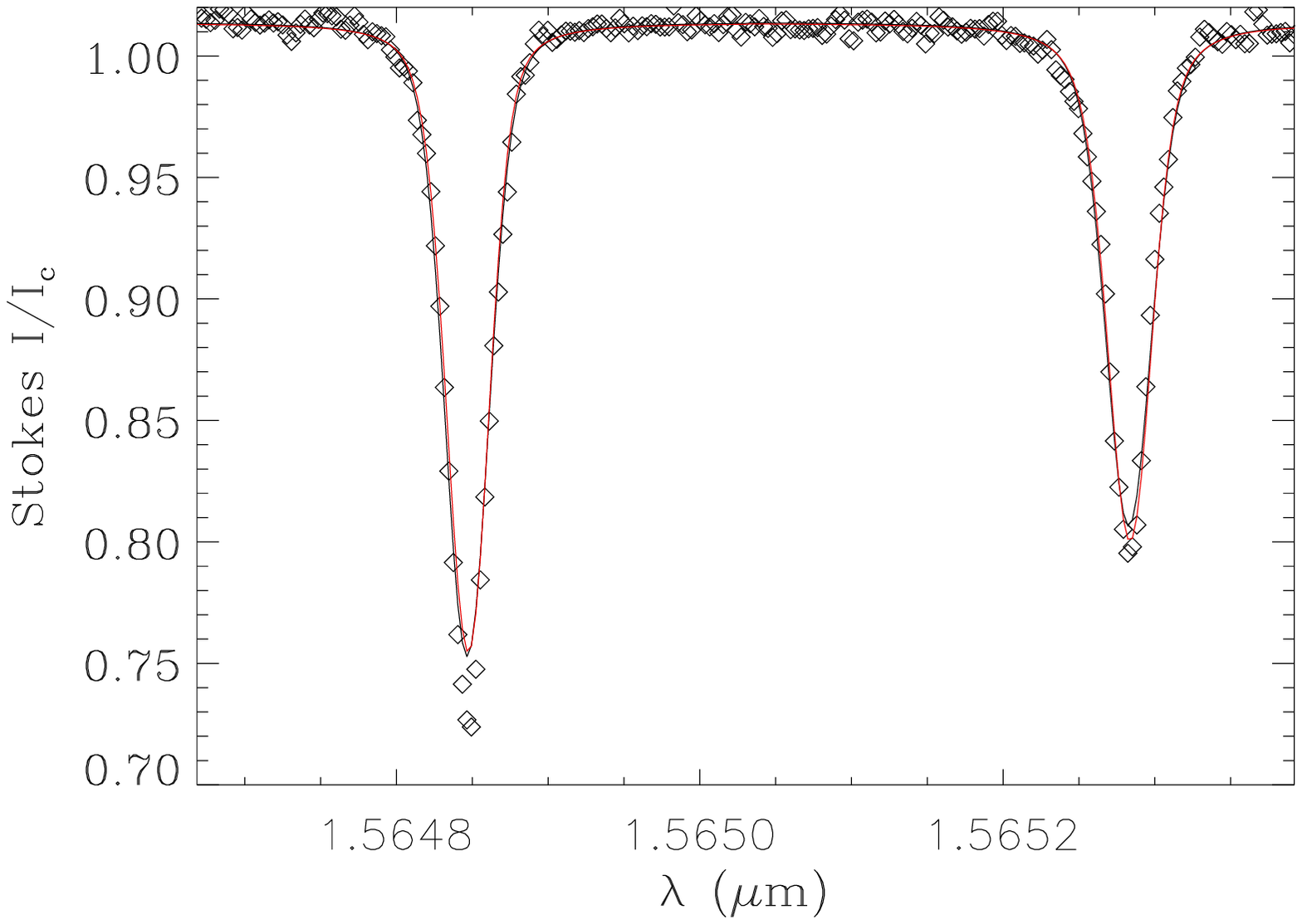}%
\includegraphics[width=8cm]{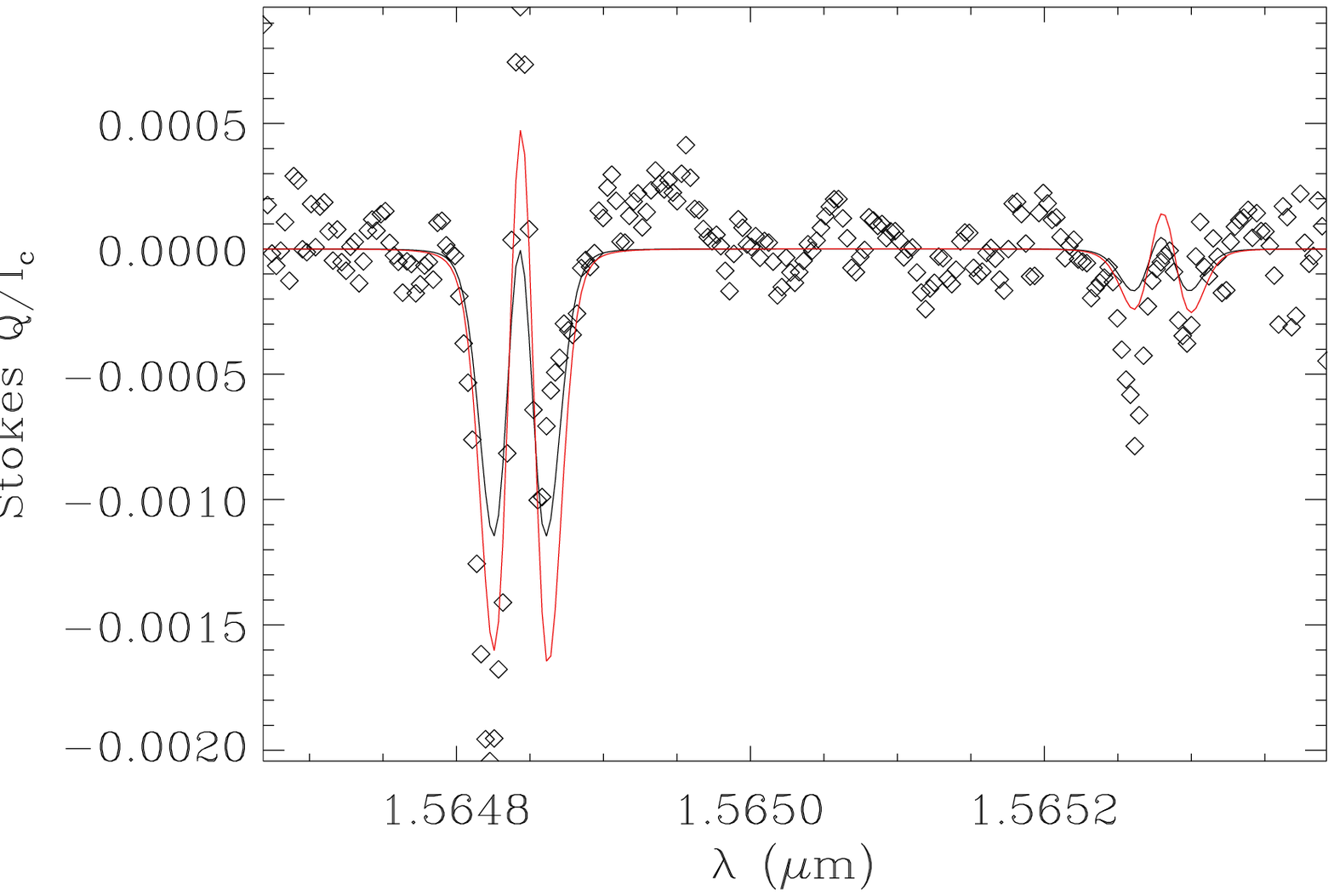}\\
\includegraphics[width=8cm]{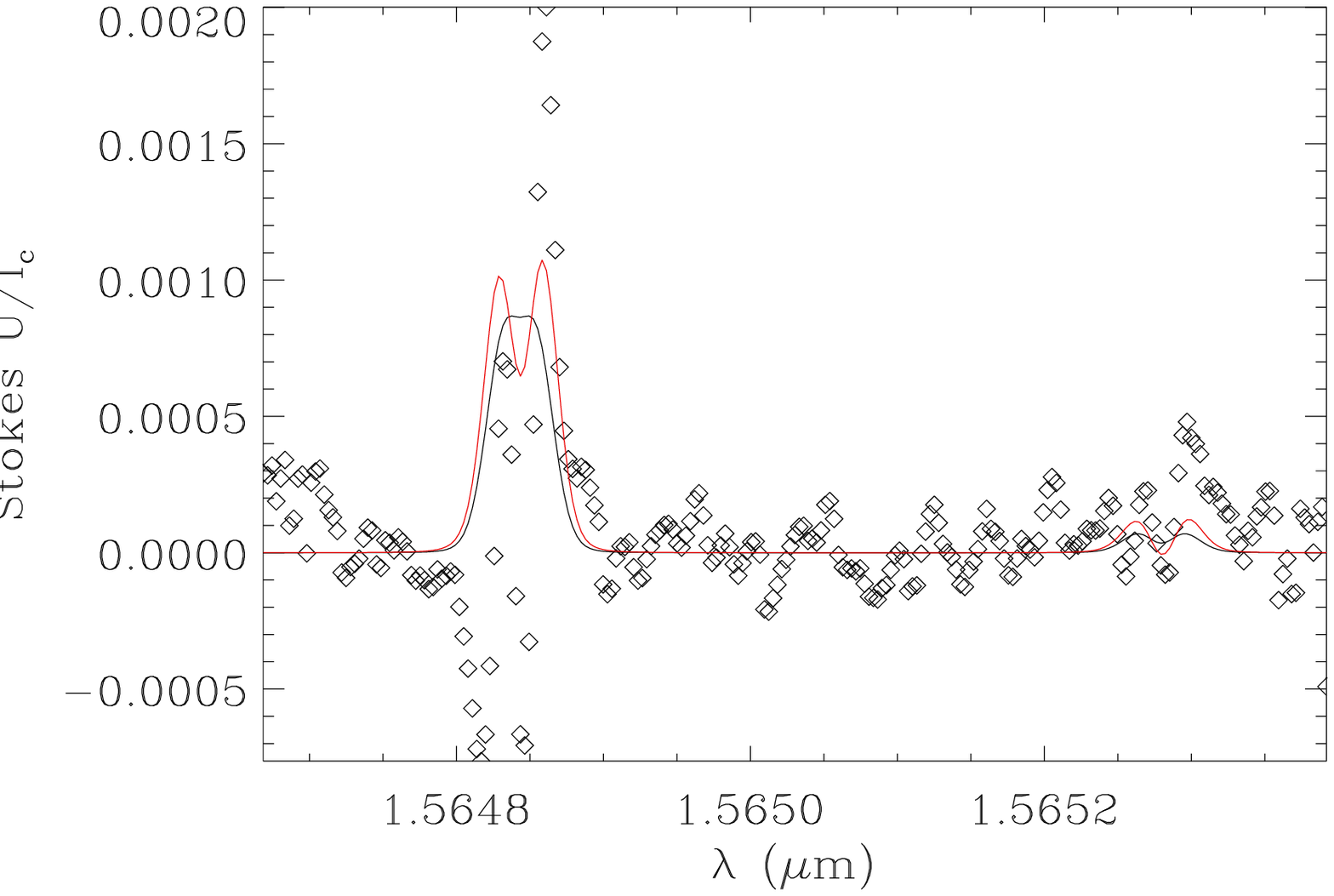}%
\includegraphics[width=8cm]{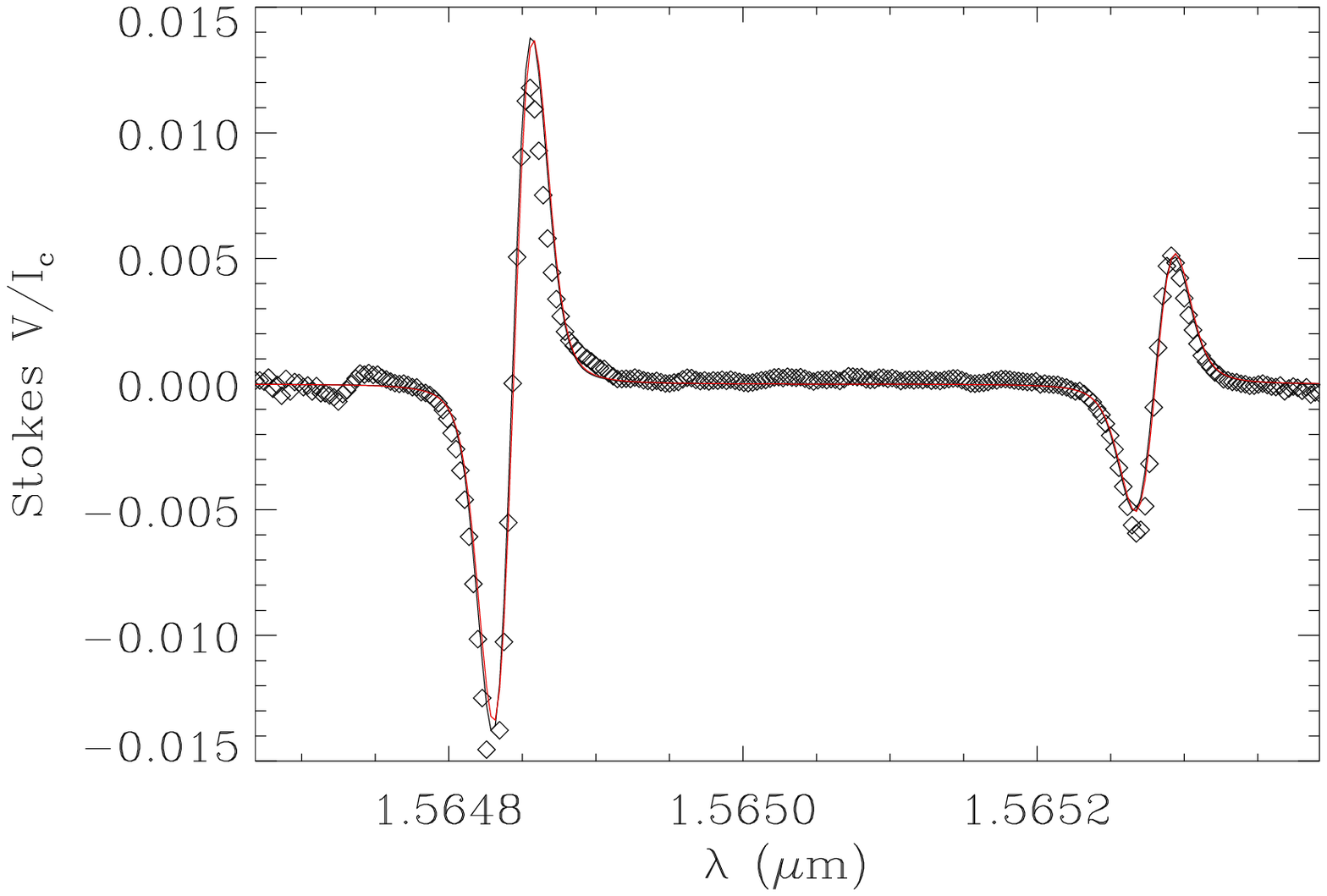}
\caption{The four upper panels represent the four Stokes profiles of a $630$ nm
observation. The four lower panels are the Stokes profiles of the co-spatial
$1.56$ $\mu$m pixel. In all the plots, diamonds represent the experimental data and 
solid lines the best fit from the two inversion procedures: the black line shows the separate inversion and red one the simultaneous one.}
\label{ejemplos}
\end{figure*}

\subsection{Asymmetries of the circular polarization}

In order to compute the asymmetries of the Stokes V profiles, we reduced the
sample by selecting only
those profiles that have two lobes in both spectral ranges. The procedure to
find the amplitude and position in wavelength of the lobes is as follows. First,
we calculated the positions where the Stokes V profile reaches zero. Then, we
computed the maximum (or minimum) of the profile between two zero-crossings.
With this procedure, the two spectral lines in the same wavelength range
(e.g.~630.15~nm and 630.25~nm) can show a different number of lobes. To select
only those lobes that are compatible in the two spectral ranges, we imposed
additionally the condition that the zero-crossings have to be the same within an error
of 1.7 km/s in the infrared data and 2.1 km/s in the visible data (equivalent to 3 pixels).
Finally, only those profiles with two lobes in both spectral
ranges were selected. With this, we ended up with 24\% of the observed area. 
We used the definition of the amplitude ($\delta a$) and area ($\delta A$) asymmetries 
of \cite{sami_84}. The amplitude asymmetry is defined as:
\begin{equation}
\delta a=\frac{(A_V)_b-(A_V)_r}{(A_V)_b+(A_V)_r},
\end{equation}
where $(A_V)_b$ and $(A_V)_b$ represent the amplitudes of the blue and red lobes
as defined before, while the area asymmetry is defined as:
\begin{equation}
\delta A=\frac{A_b-A_r}{A_b+A_r},
\end{equation}
where $A_b$ and $A_r$ are the absolute values of the area of the blue and red
lobes, respectively. The boundaries of the integral of each lobe are different
for each profile. We choose the initial point of the integration of the blue
lobe as the position in the blue wing closest to the lobe which has an amplitude lower than
$3\times 10^{-5}$ I$_\mathrm{c}$ and the ending point the zero-crossing of the
profile. This last wavelength is the lower limit of the integration of the red lobe and the
final one is again the wavelength point in the red wing closest to the profile having an amplitude lower
than $3\times 10^{-5}$ I$_\mathrm{c}$. 

Figure \ref{asim_tip_polis} shows the histograms of the amplitude and area
asymmetry. The amplitude asymmetry distributions for the visible and infrared data 
peak at 0.15. Both are similar in shape, but the one corresponding to the visible lines is slightly 
broader. The infrared distribution is very similar to that of
\cite{khomenko_03}, who also found a peak at 0.15. The histograms of the area
asymmetries are more symmetric than those of the amplitude asymmetries and are
both centered at 0. Again, the visible distribution is broader than the infrared
one. For comparison, the infrared histogram shown in \cite{khomenko_03} peaks at
0.07 and seems to be narrower. Positive amplitude asymmetries and zero area 
asymmetries mean that our Stokes profiles have 
a higher amplitude and narrow blue lobe and a small amplitude and broad red
lobe. This kind of 
asymmetric profiles are typical of active regions like plages and network
\citep{sami_84}.

\begin{figure*}
\centering
\includegraphics[width=9cm]{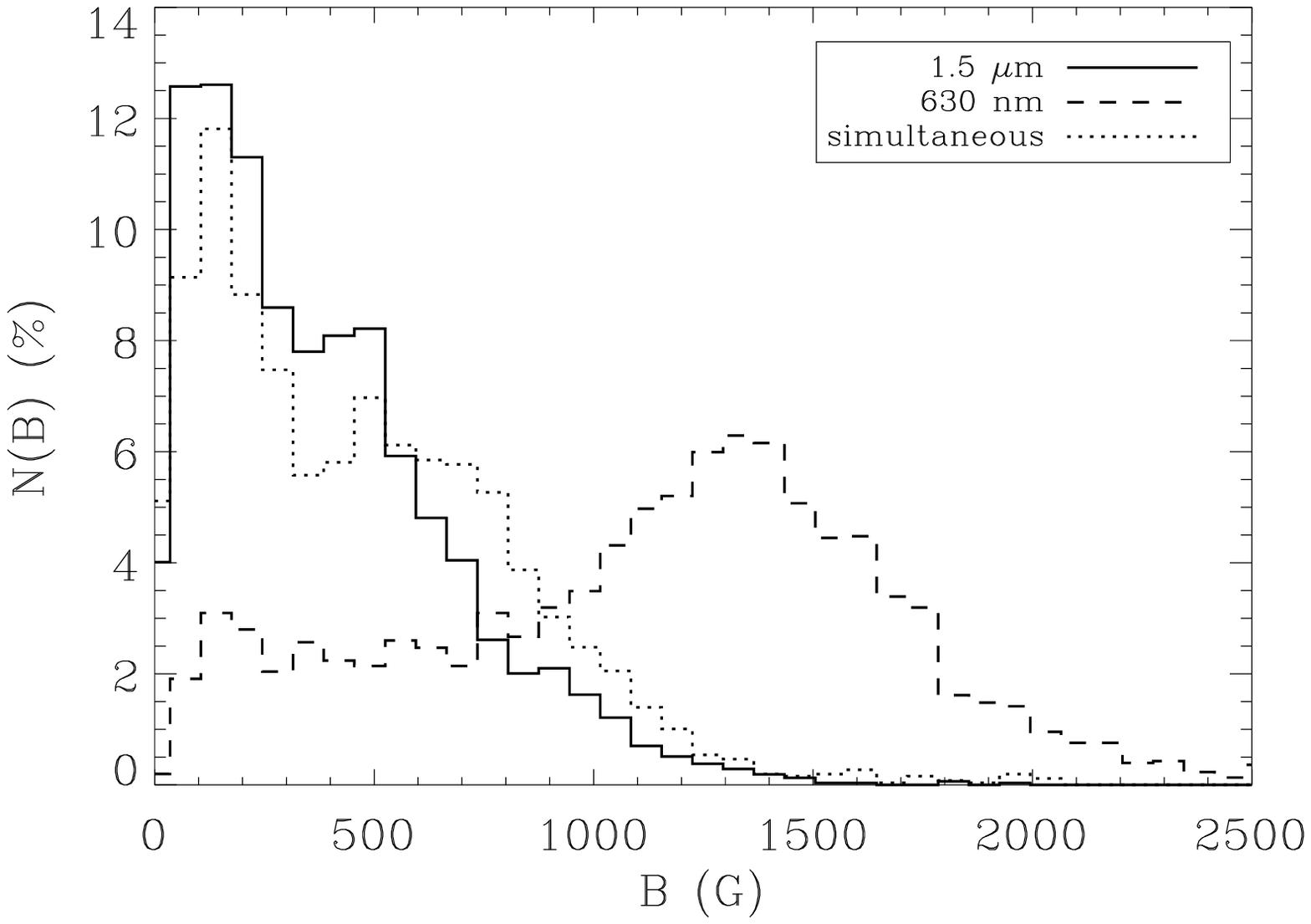}
\includegraphics[width=9cm]{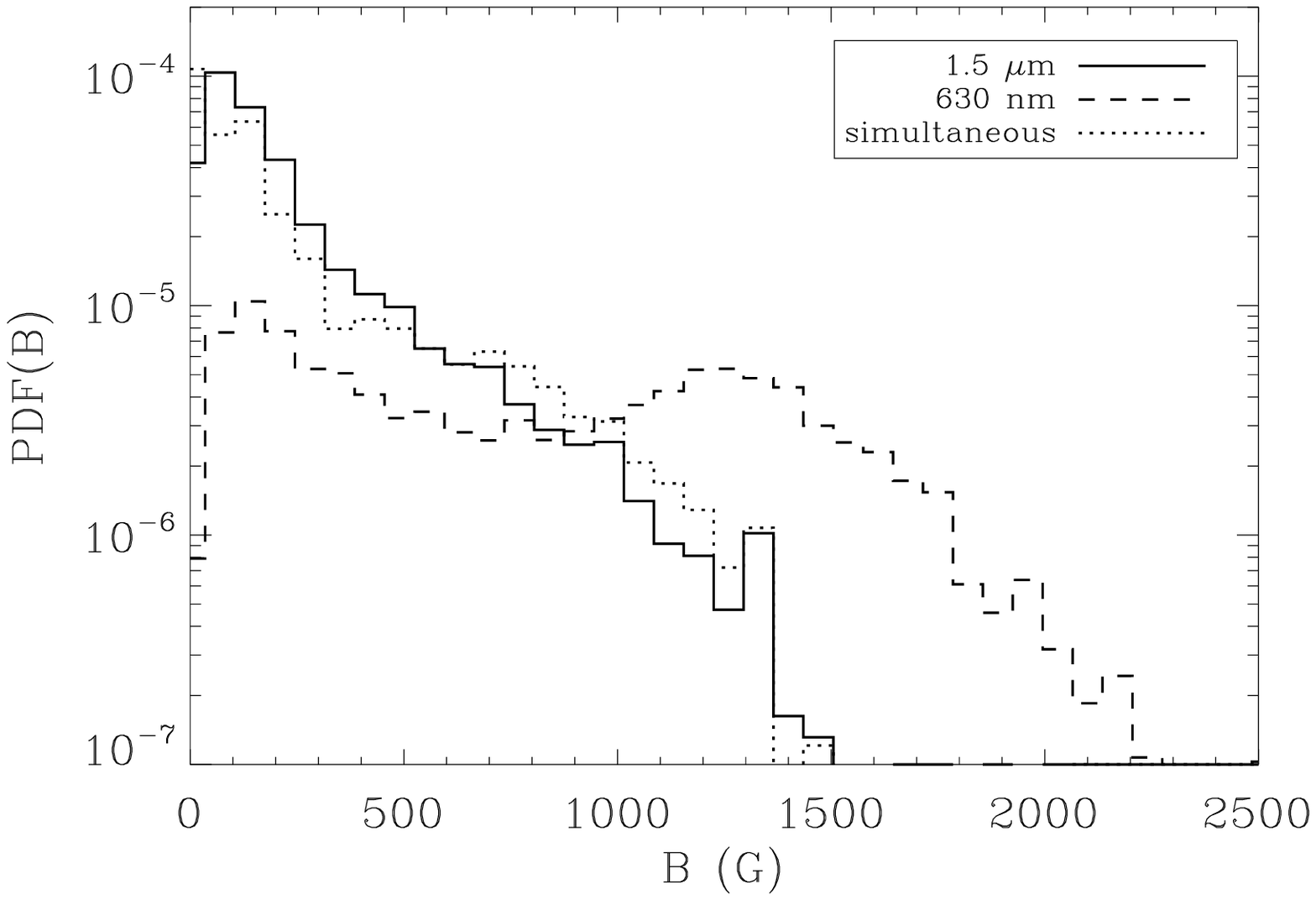}
\caption{Magnetic field strength distributions at internetwork regions recovered
from 
the separate inversion of the \hbox{$1.56$ $\mu$m} pair of lines (solid line)
and the $630$ nm ones (dashed line) and from the simultaneous inversion of the
two data sets (dotted line).
The left panel shows the histogram of the magnetic field strength. The right panel 
represents the PDF. Both representations have been plotted to easily compare
with previous works. Note that the PDF takes into account the filling factor.}
\label{pdf_separat}
\end{figure*}

\begin{figure*}
\centering
\includegraphics[width=9cm]{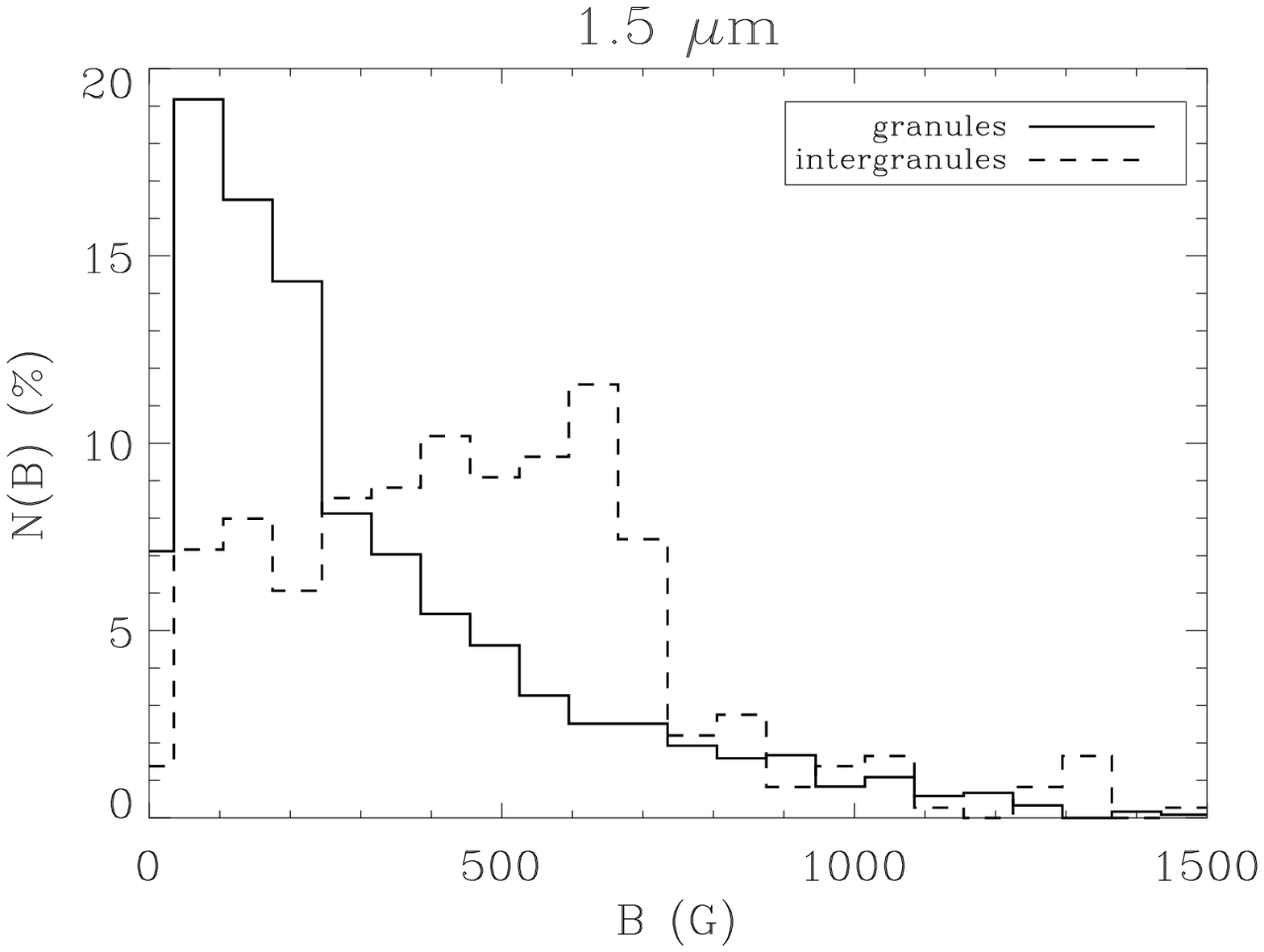}
\includegraphics[width=9cm]{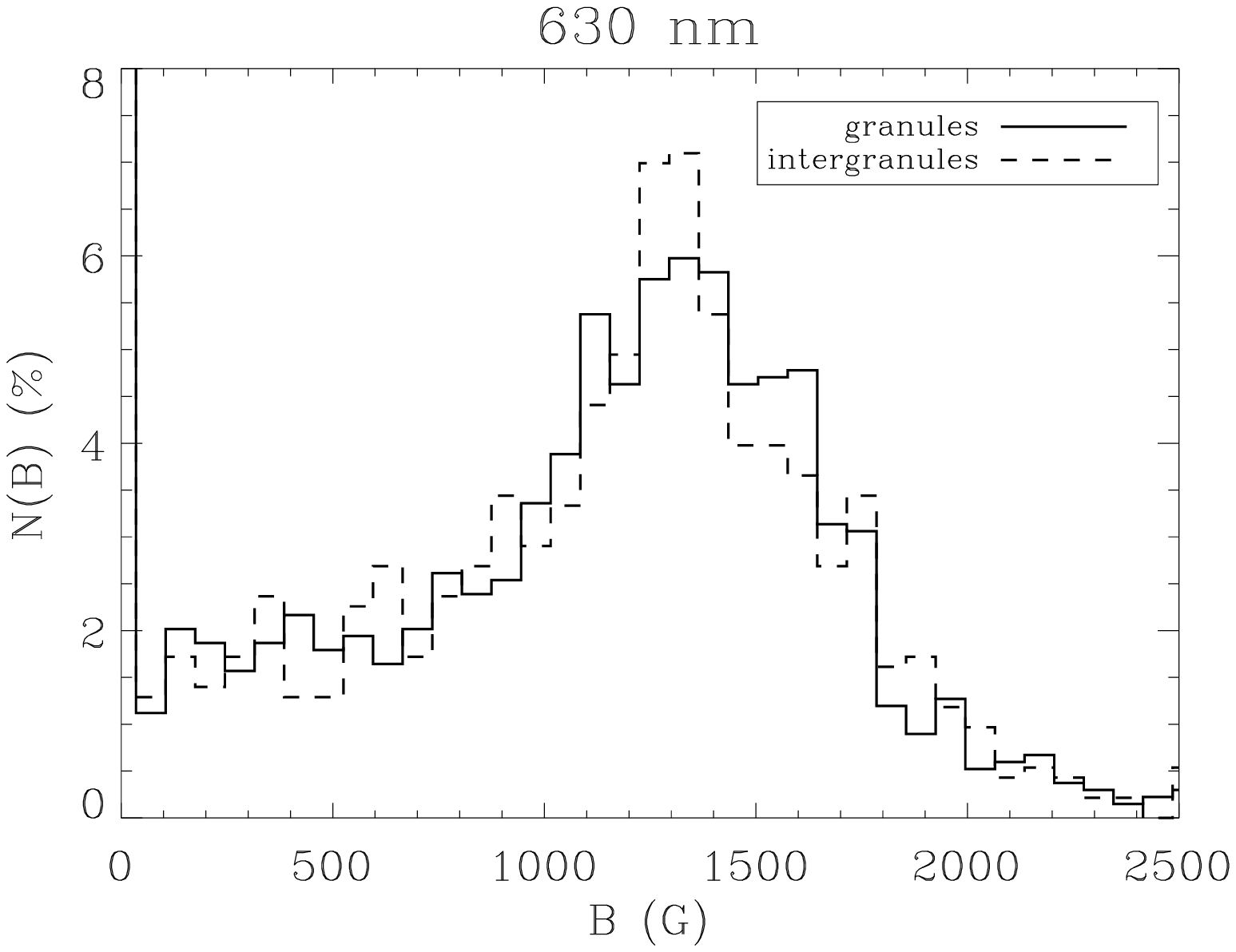}
\caption{Granular (solid line) and intergranular (dashed line) magnetic field
strength distributions 
recovered from the separate $1.56$ $\mu$m inversion (left plot) and from the
$630$ nm one (right plot).}
\label{pdf_gran_intergran_separat}
\end{figure*}

\section{Inversion of the full Stokes vector}

Inversion procedures are very powerful diagnostic techniques that allow us to
retrieve the physical conditions of the solar atmosphere from the information
contained in the Stokes vector. However, the smaller the signal-to-noise ratio
the smaller the information contained in the spectra. This was our main reason
for using the PCA to reduce the noise contribution. We used the
SIR code \citep{basilio_92} to carry out the inversions of our data sets. We
adopted a two-component model to reproduce the observed signals in each pixel: a
magnetic atmosphere covering some fraction of it and a field-free one filling
the rest of the surface. This modeling only allows us to reproduce regular
anti-symmetric V profiles. Consequently, we only used two-lobed V profiles with a
signal-to-noise ratio above $10\sigma$ in both spectral ranges. 

The free parameters of the inversion for the field-free atmosphere are: 
the temperature height profile (with a maximum of 5 nodes), the 
line-of-sight (LOS) velocity height profile (with a maximum of 3 nodes) and the
microturbulent velocity. Concerning the magnetic component, the parameters are:
the temperature height profile (with a maximum of 5 nodes), the microturbulent and the
LOS velocity, the magnetic field strength, the inclination of the magnetic field
vector with respect to the LOS and the field azimuth. The magnetic field vector and the
LOS velocity are set constant with height. One single macroturbulent velocity 
was applied to the synthetic profiles. The filling factor of the magnetic component 
is also a parameter of the inversion.

As a first step, we performed separate inversions of the \hbox{$1.56$ $\mu$m} and \hbox{630 nm}
data. As will be demonstrated below, since the information contained in both
spectral ranges is compatible, we also carried out an inversion of both spectral
ranges simultaneously.

The assumption of magnetic field properties and LOS velocities constant with
height prevents the recovery of the asymmetries of the Stokes V profiles. Thus,
profiles with strong asymmetries cannot be well reproduced, although other
properties like the amplitude ratio between the two lines (for each pair of
lines) or the width of the Stokes V lobes can be correctly fitted.
Figure \ref{ejemplos} shows examples of observed and best-fit
profiles for the \hbox{$1.56$ $\mu$m} and \hbox{630~nm} lines for the separate (black line)
and the simultaneous inversion (red line). In both cases, Stokes
I and V are correctly reproduced. The observed linear polarization signals are
extremely low and show some crosstalk contamination (mainly at 630 nm). In spite of this, the
shapes and the amplitudes of the profiles are nicely reproduced.


\begin{figure*}
\includegraphics[width=9cm]{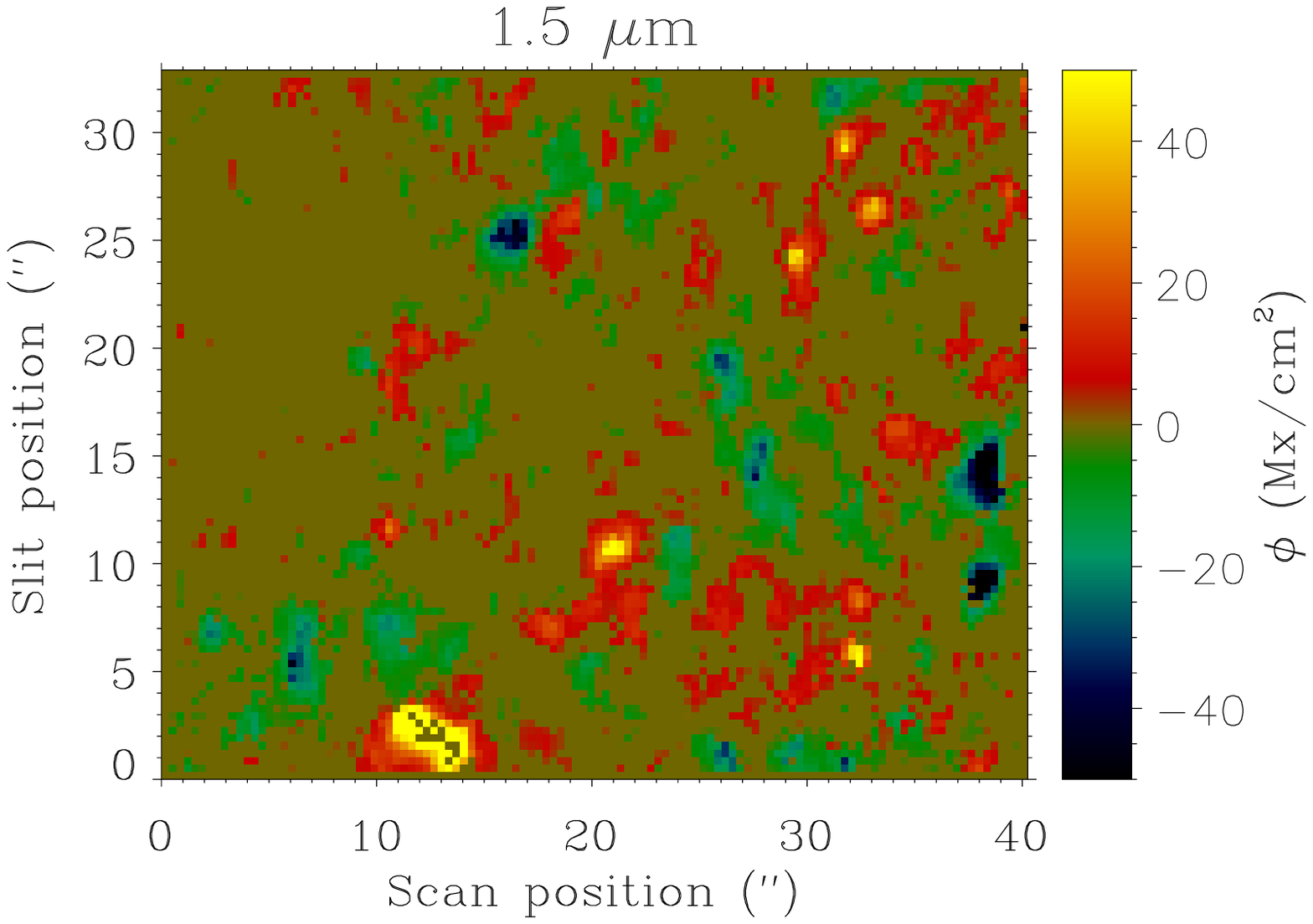}
\hspace{0.3cm}
\includegraphics[width=9cm]{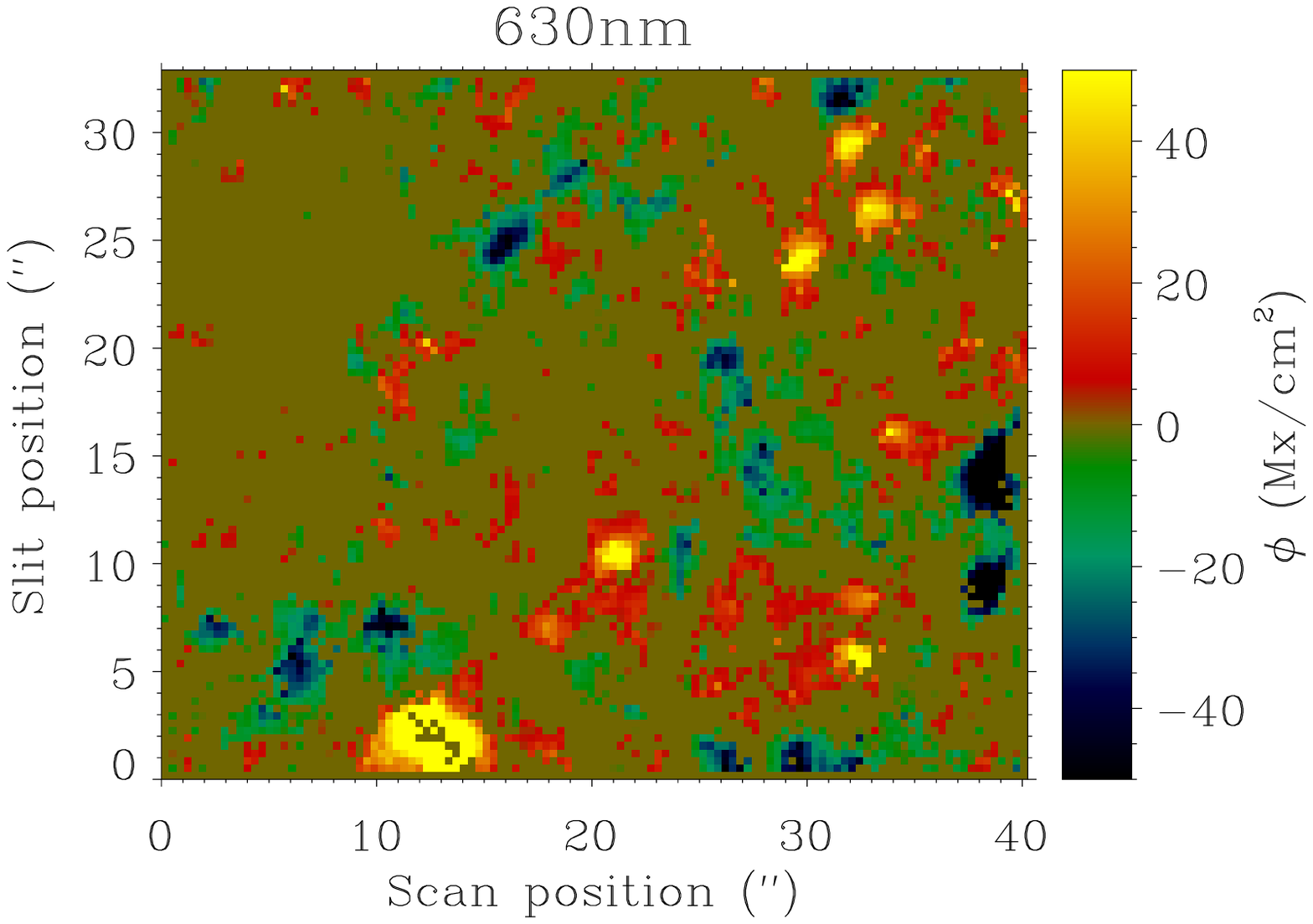}
\caption{Magnetic flux density maps retrieved from the separate inversions of
the $1.56$ $\mu$m and $630$ nm data. 
Both maps have been saturated to the range -50 and 50 Mx/cm$^2$.}
\label{mapas_flujo}
\end{figure*}

\section{Results}


\subsection{Magnetic field strength distributions from independent inversions of
the 1.5 $\mu$m and 630 nm data sets}

Figure \ref{pdf_separat} shows the magnetic field strength distributions
retrieved from the separate inversion of the infrared and visible data sets
(solid and dashed lines, respectively). The left panel represents the histogram of
the magnetic field strength for a direct comparison with \cite{khomenko_03}.
The right panel represents the probability density function (PDF) for a direct
comparison with \cite{jorge_ita_03}. The PDF is the probability of finding a
particular range of magnetic field strengths between $B$ and $B+dB$ in the field of view. It has been
computed from the inversion results as:
\begin{equation}
PDF(B,B+\Delta B)= \frac{\sum_i f(B_i,B_i+\Delta B_i)}{N \Delta B},
\end{equation}
where $f(B_i,B_i+\Delta B_i)$ is the value of the magnetic filling factor for
the magnetic field strengths that are between $B_i$ and $B_i+\Delta B_i$. The symbol $N$ 
represents the total number of profiles in the field of view and $\Delta B$ denotes 
the bin size. Note that the PDF takes into account the filling factor of the magnetic elements. 

The $1.56$ $\mu$m inversions reveal that the majority of magnetic fields at
the internetwork have field strengths well below kG. This has been already
pointed out by other works using the same pair of lines \citep{lin_95,lin_99,khomenko_03}. 
It must be noted that the shape of the infrared distribution resembles that 
presented by \cite{khomenko_03}. On the contrary, the field strength
distribution obtained from the visible lines has a peak at kG fields. This 
result is in agreement with the
works of \cite{hector_02}, \cite{ita_jorge_03}, \cite{jorge_ita_03},
\cite{lites_04} and \cite{ita_06}. 
The approximate height of formation of the infrared pair of lines is deeper than for
the visible ones. However, this difference in height ($\sim 200$ km) is not big enough to explain
the extremely different values of the field strength. At this point, the
magnetic field strength recovered separately from the $1.56$ $\mu$m and $630$ nm
inversions are apparently incompatible.

In order to check the reliability of the inferred magnetic field distributions,
we study the distributions of granular and intergranular regions. The spatial
resolution in our data was good enough to show a correlation between the
continuum intensity and the bulk velocity, in the sense that dark areas in our
data statistically correspond to downflows (intergranular lanes) and bright
areas to upflows (granules). We select as granules those pixels where the continuum
intensity was larger than the mean continuum intensity and intergranules those
which present smaller values. Figure \ref{pdf_gran_intergran_separat} shows
the magnetic field strength distribution for granular and intergranular regions.
In the infrared we can clearly distinguish two different distributions. 
In the granular cells, the fields are intrinsically weaker and have a
distribution with a tail with an exponential decay. In the
intergranular lanes, the distribution is centered at values near the
equipartition field on the photosphere ($\sim 500$ G). This result is expected
from magnetohydrodynamical considerations and was already pointed out by
\cite{lin_99} and \cite{khomenko_03}. The scenario is very different in the
visible data: we found the same distribution in granules and intergranules. This
strange behaviour made us think that the inversion of the $630$ nm
lines alone might not be reliable. 

The magnetic filling factors inferred from the analysis of the $1.56$ $\mu$m data
is $1-2$\% for magnetic field strengths 
larger than \hbox{$300$ G}. Smaller field strengths are in the weak field
regime and the magnetic field strength can not be separated from the
filling factor. In the visible, the filling factor is $0.5-1$\% for field
strengths higher than approximately $500$ G.

\subsection{Compatibility of the observations}

In this section, we present arguments about the compatibility of our data taken at 
$1.56$ $\mu$m and \hbox{630 nm} to be sure that the simultaneous inversion of both 
profiles at different wavelengths is really trustful. Some authors working mainly with the $630$ nm
lines suggested that both spectral ranges carry different information about the
magnetic field \citep{hector03}. These conclusions are based 
on the systematical presence of opposite polarities of co-spatial circular polarization profiles
at both wavelengths \citep{hector_04}. \cite{jorge_ita_03} analysed simultaneous observations in both
spectral ranges, discovering the presence of opposite polarities in $25\%$ of
their selected profiles. Another ingredient they point out is the higher flux density
inferred from visible wavelengths as compared to the one recovered from the
infrared. \cite{jorge_ita_03} measured an unsigned flux density at \hbox{$630$ nm} that
was almost two times the one at $1.56$ $\mu$m. 

\begin{figure}
\hspace{-0.3cm}
\includegraphics[width=9.5cm]{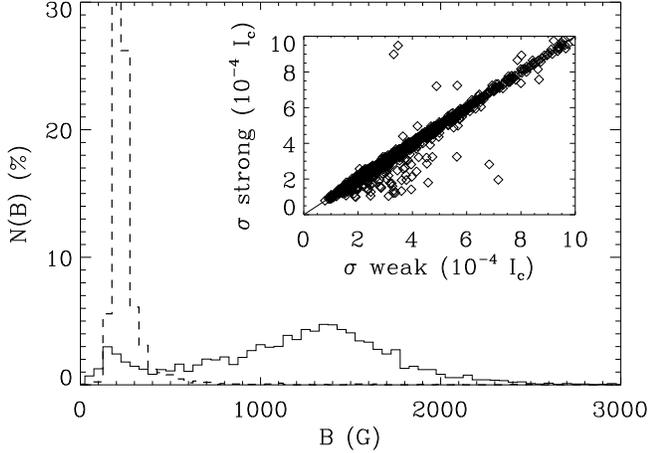} 
\caption{Magnetic field strength distributions recovered from the inversion of the 630 nm
data set with a kG (solid line) and 
a \hbox{200 G} (dashed line) magnetic field strength initialization. The inset window
represents the goodness of the fit in terms of the standard deviation of the
difference between the observed and fitted Stokes V profiles.}
\label{pdf_fort_debil}
\end{figure}

In our simultaneous observations, opposite polarities 
are not a common occurrence, amounting only to $3.6\%$ of our two-lobed selected
profiles. Figure \ref{mapas_flujo} shows the spatial distribution of the
magnetic flux density recovered from the separate analysis of the infrared and visible
lines, defined as:
\begin{equation}
\phi=fB\cos{\theta},
\end{equation}
being $\theta$ the inclination of the magnetic field vector with respect to the
LOS. Both infrared and visible lines measure the same magnetic flux density, as
shown by the strong correlation between both maps. Only those
areas of more intense magnetic flux density are slightly larger in the visible
than in the infrared. Not only the value of the magnetic flux density is compatible 
but also the polarities are the same in the same pixel in both spectral ranges. Consequently, the
conclusion can be reached that both 
spectral ranges are tracing the very same magnetic features in the solar surface. The following
question arises: if the observations are compatible in both spectral ranges, why does the magnetic field
strength distribution at both spectral ranges seem to be incompatible?

\begin{figure}[!t]
\hspace{-0.3cm}
\includegraphics[width=9.5cm]{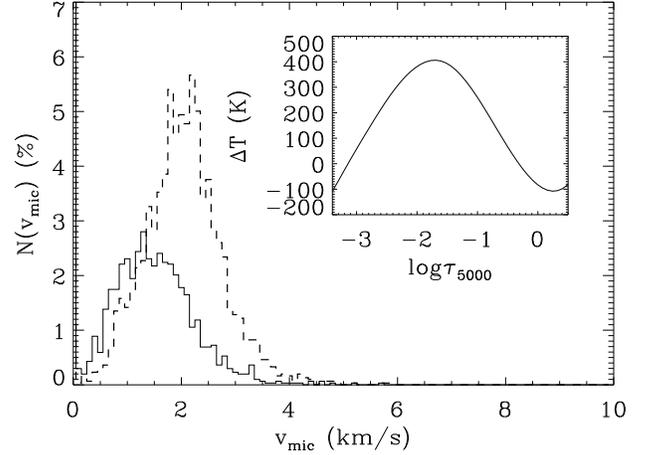} 
\caption{Magnetic microturbulent velocity distributions recovered from the
inversion of the 630 nm data set with a kG (solid line) and a \hbox{200 G} (dashed
line) magnetic field strength initialization. The inset window represents the
difference between the mean temperatures of the magnetic component inferred from
the weak and strong initializations.}
\label{vmic_t_fort_debil}
\end{figure}

\cite{marian_06} showed that the 630 nm lines do not carry binding information
about the magnetic field strength at internetwork areas. They showed that with
typical present internetwork observations (magnetic flux density \hbox{$\sim$ 10
Mx/cm$^2$} and a noise level of $6.5\times 10^{-5}$ I$_\mathrm{c}$) the magnetic
field strength can not be reliably recovered and separated from the thermodynamic parameters
of the atmosphere. 
In order to check the reliability of the retrieved magnetic field distribution
from the independent $630$ nm inversions, we perform two different inversions
with two different initializations of the code: one with a kG field (the one
showed in Fig. \ref{pdf_separat}) and another one with an initial sub-kG field
strength (\hbox{$200$ G}). 
Figure \ref{pdf_fort_debil} shows the two inferred magnetic field distributions
from the two different inversion procedures. The one starting with a kG field
results in a field strength distribution with a predominance of kG magnetic fields, 
while the one initialized in the sub-kG regime is radically different, presenting a 
peak around $200$ G. The inset window shows the mismatch (in terms of the standard
deviation, $\sigma$) between the observational Stokes V profiles and the fits
for both initializations.
Although the whole Stokes vector is well reproduced, we have only taken 
Stokes V into account to avoid our results to be biased by the intensity
profile. All the points remain in 
the diagonal, meaning that indistinguishable profiles are obtained from both
different inversions. Figure \ref{vmic_t_fort_debil} shows the microturbulent
velocity histograms of the magnetic component retrieved from the inversion of
the 630 nm data set with the weak and strong initializations. As can be seen, the
initialization with a weak field gives rise to larger values of the microturbulent 
velocity. This is used by the inversion code to broaden
the profiles since the magnetic field is not strong enough. The inset window in
the same figure shows the difference between the mean temperatures of the
magnetic component retrieved from the weak and strong initializations. The
presence of different temperatures at the (different) heights of formation of the $630.15$ and
$630.25$ nm lines are used to modify the relative amplitudes of their Stokes V profiles in
order to compensate for the weak value of the field strength. Consequently, since 
the magnetic field strength cannot be disentangled from the thermodynamics, one
cannot discriminate between the two recovered magnetic field strength distributions shown
in Fig. \ref{pdf_fort_debil}.

According to our results, the claim by \cite{hector03} that there is an observational bias produced
by the difference in wavelength of the two spectral regions does not sustain until
reliable results are found for visible lines. We have been able to
obtain sub-kG or kG fields from the same data set in the visible lines.
Recently, Ram\' irez V\'elez, L\'opez Ariste \& Semel (private comunication) have used the Mn\,{\sc i} spectral line at $553.7$ nm 
and find results that are in agreement with those obtained from the $1.56$ $\mu$m, 
with a distribution of magnetic field strengths in the internetwork well below the kG regime.
The observations at $1.56$ $\mu$m and 630 nm are compatible and a simultaneous inversion based on a single
magnetic component does not lead to contradictions and appears to be sufficient.


\begin{figure*}[!t]
\hspace{-0.6cm}
\includegraphics[width=9cm]{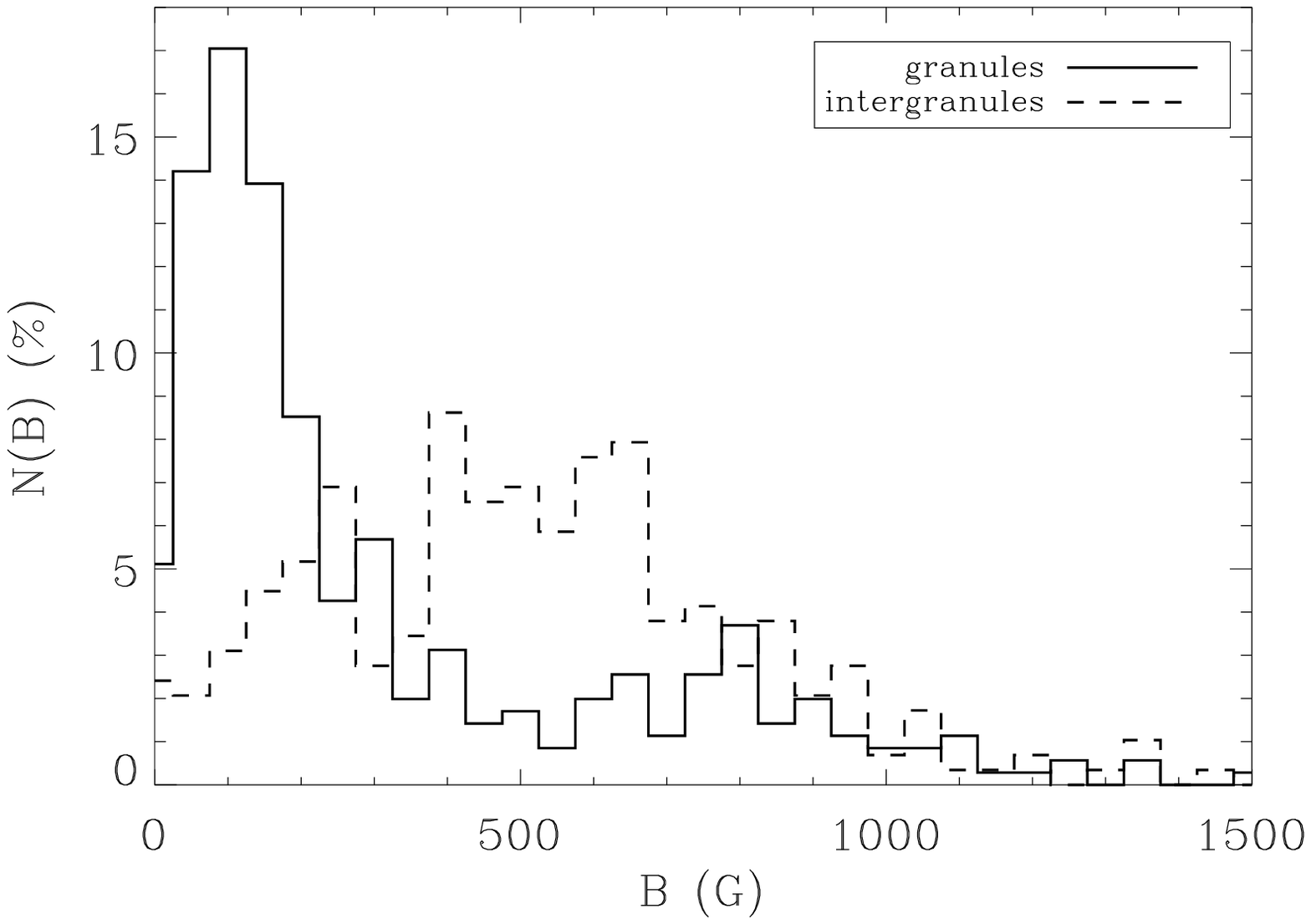}
\hspace{0.2cm}
\includegraphics[width=9cm]{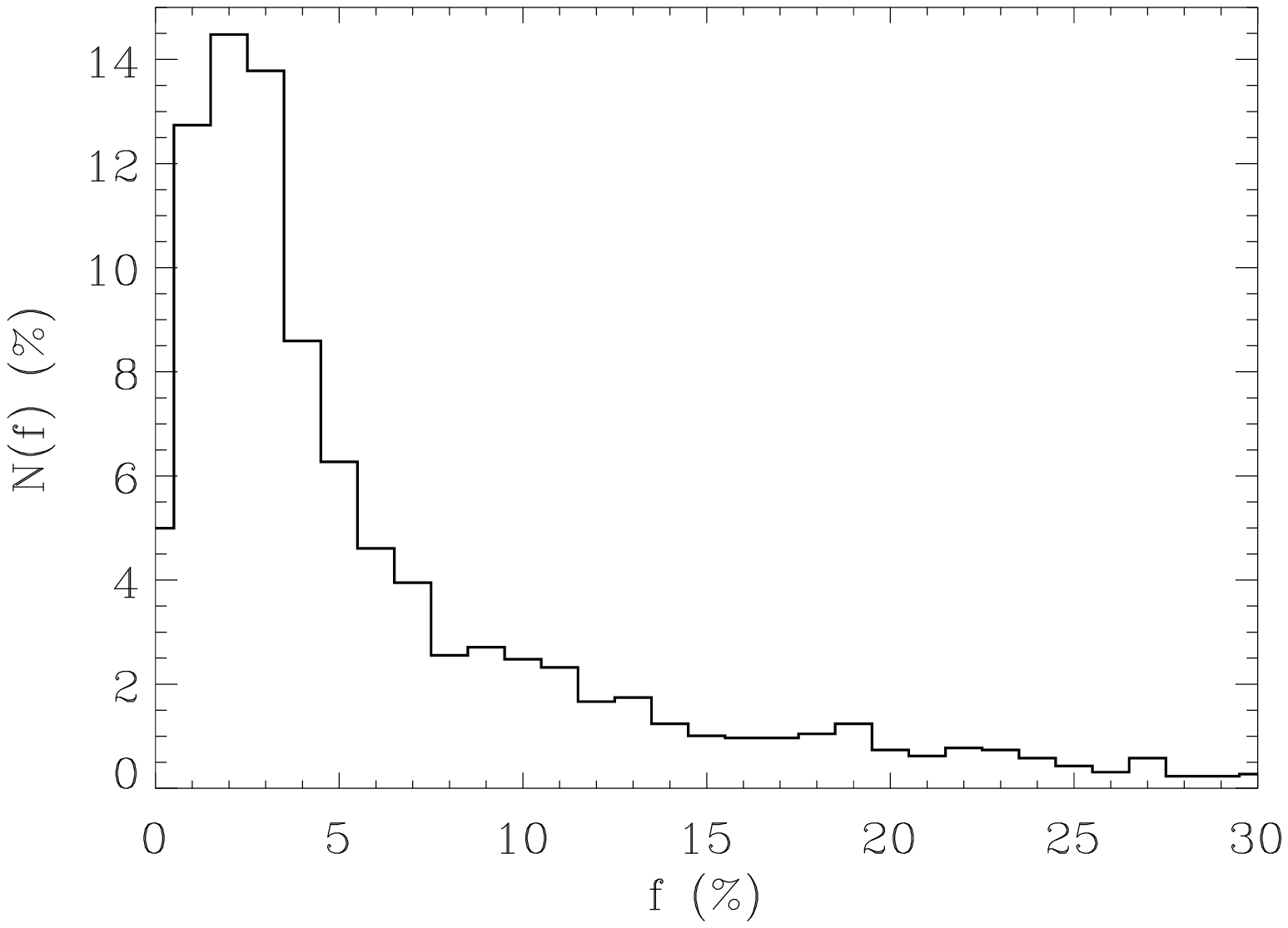}
\caption{Left panel: granular (solid line) and intergranular (dashed line) magnetic field 
strength distributions recovered from the simultaneous $1.56$ $\mu$m and $630$ nm
inversions. 
The dotted line is the total distribution. Right panel: distribution of the magnetic
filling factor.}
\label{pdf_gran_intergran_sim}
\end{figure*}

\begin{figure*}
\includegraphics[width=9cm]{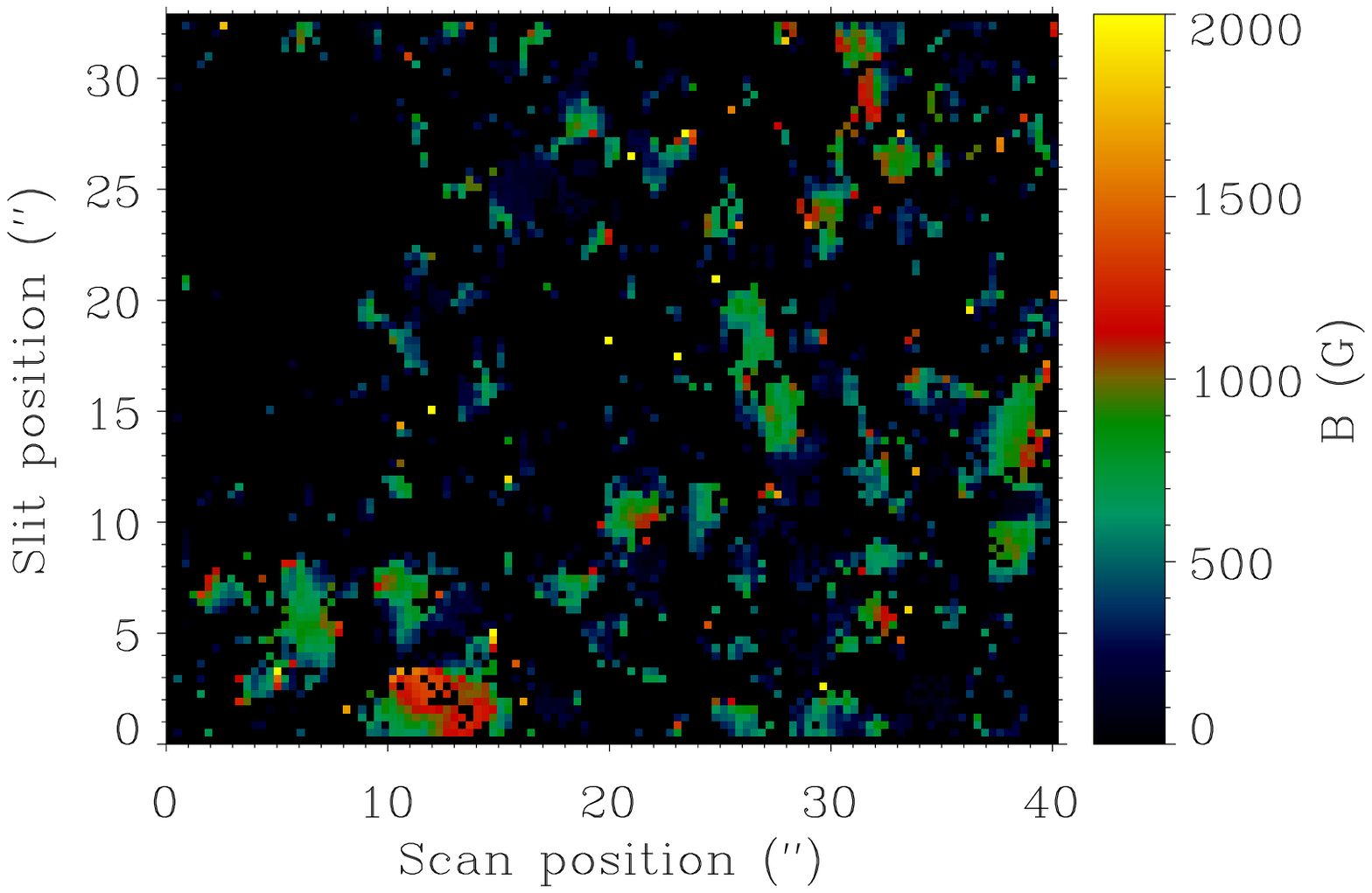}
\hspace{0.3cm}
\includegraphics[width=9cm]{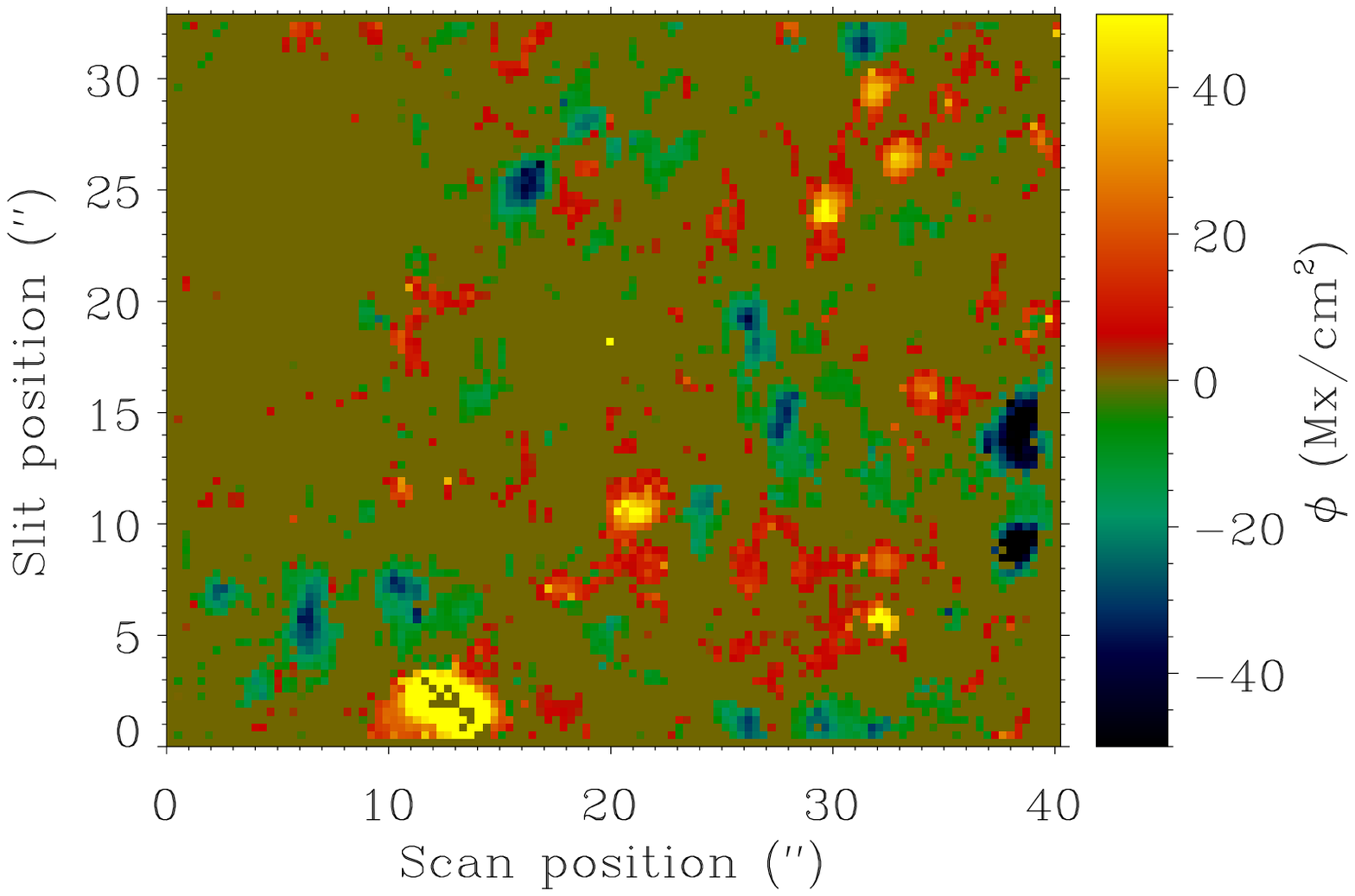}
\caption{The left panel shows the spatial distribution of the magnetic field
strength retrieved from the simultaneous inversions of the infrared and visible
data sets. the right panel represents the map of the magnetic flux density. All
points set to zero were not analyzed.}
\label{mapa_b_flujo_sim}
\end{figure*}

\begin{figure*}
\includegraphics[width=9cm]{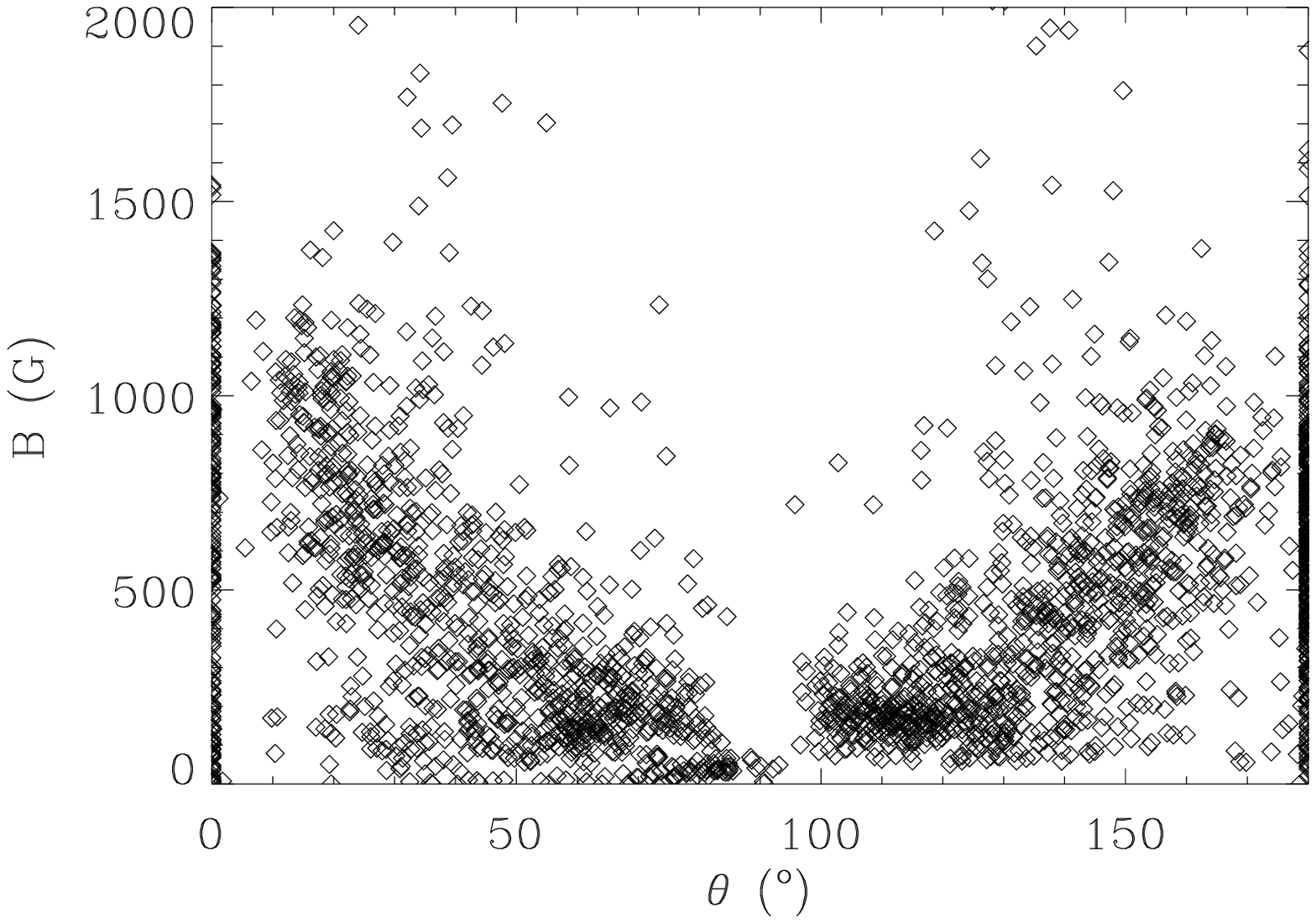}
\includegraphics[width=9cm]{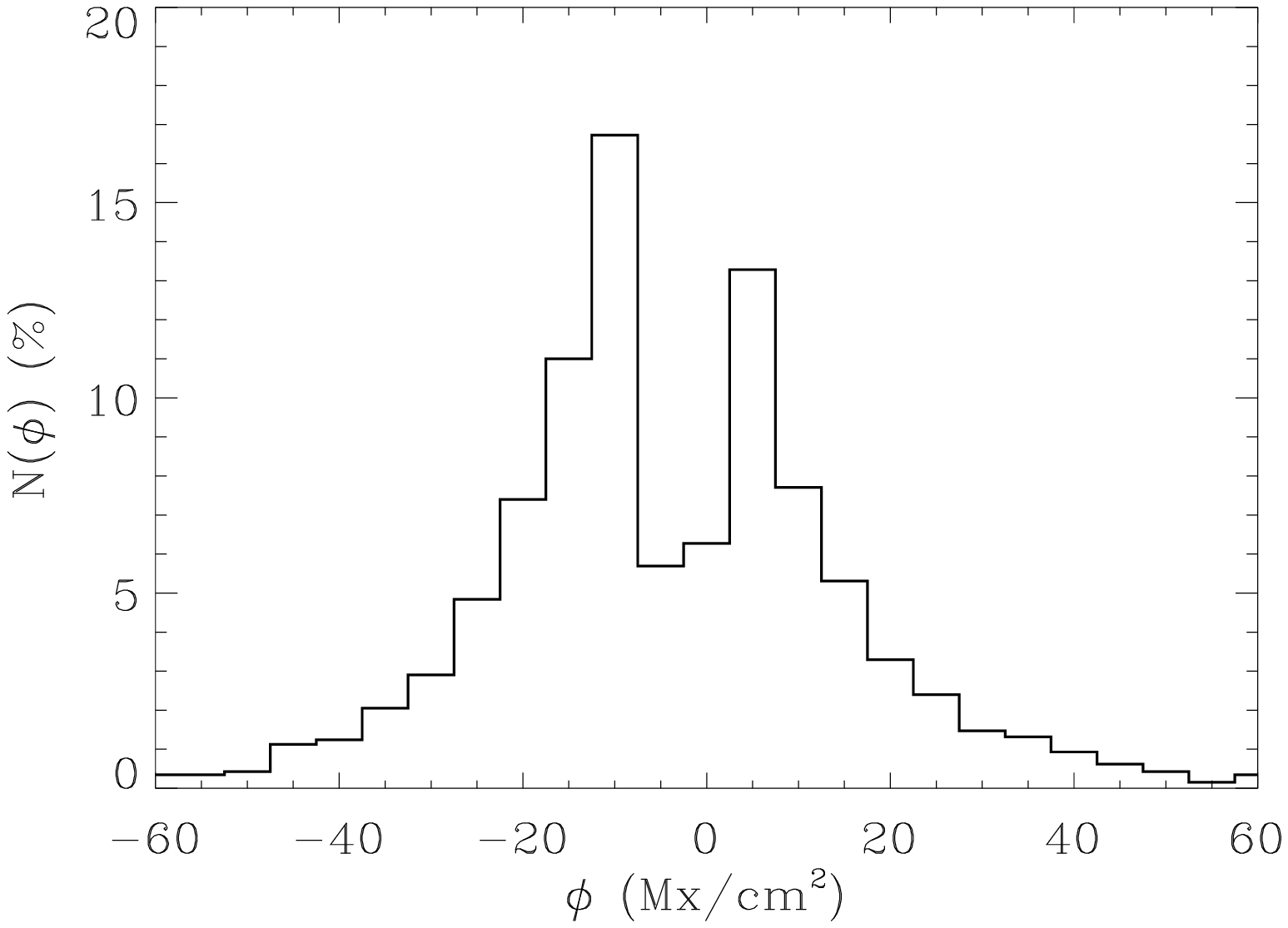}
\caption{Left panel: magnetic field strength versus the inclination with respect to
the LOS. Right panel: histogram of the magnetic flux density. All magnitudes have been  recovered from the simultaneous inversion of the infrared and
visible data sets.}
\label{theta_camp}
\end{figure*}

\begin{figure*}
\includegraphics[width=9cm]{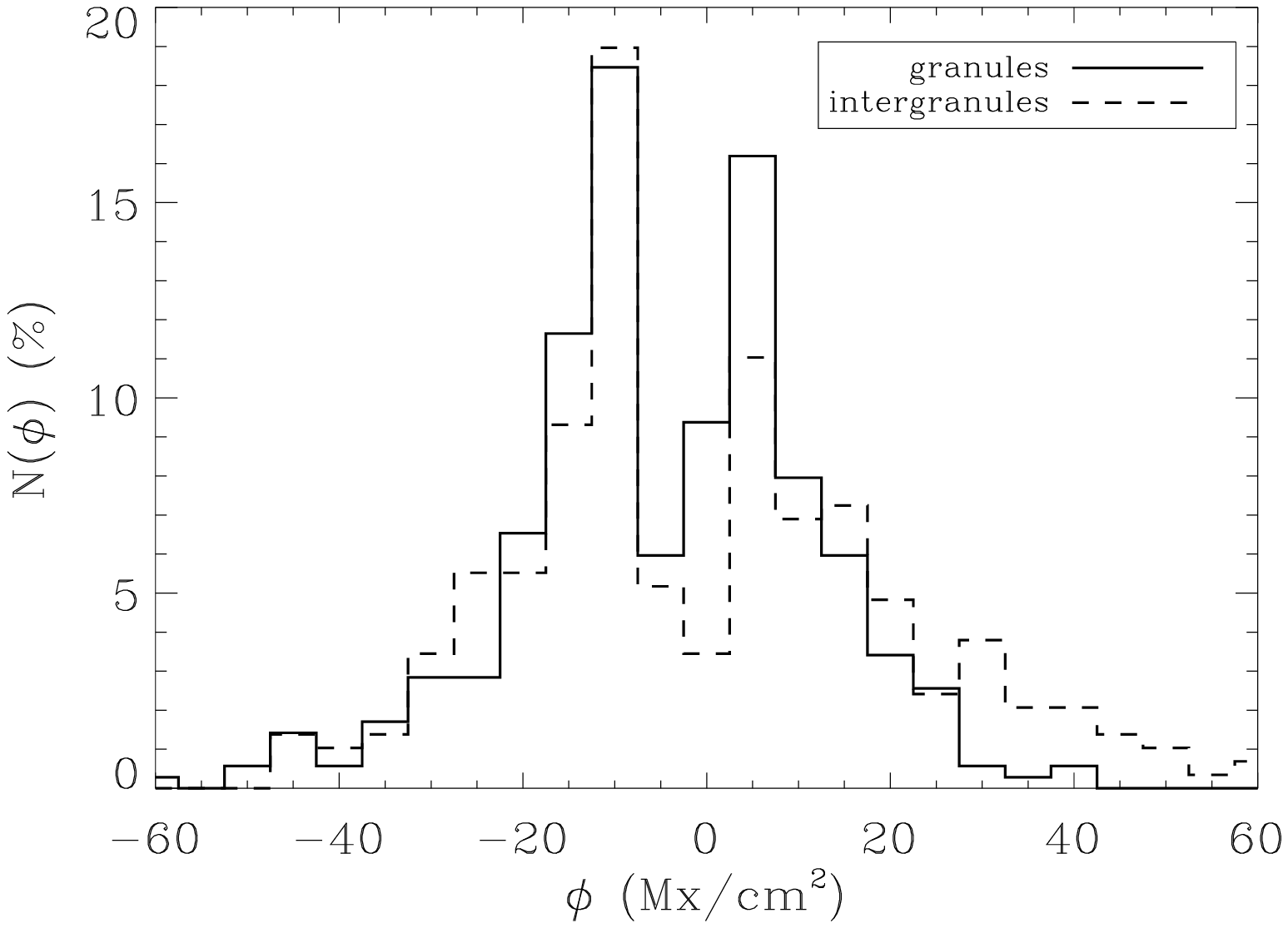}
\includegraphics[width=9cm]{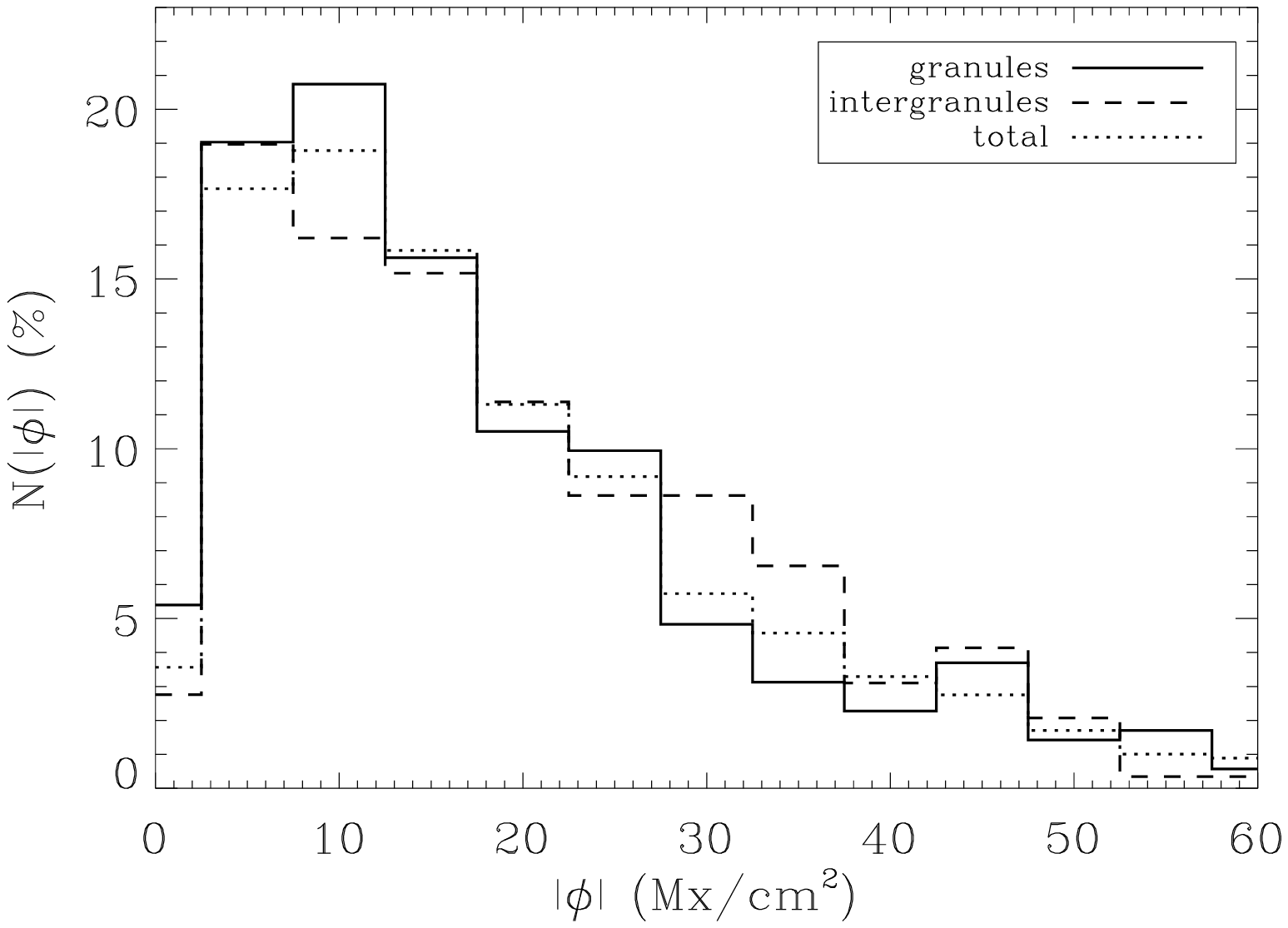}
\caption{Left panel: magnetic flux density distributions of granules (solid line) and
intergranules 
(dashed line). Right panel: unsigned magnetic flux density distributions of granules (solid line) and intergranules (dashed line). The dotted line is the total unsigned
magnetic flux distribution. All magnitudes have been  recovered from the simultaneous inversion of the infrared and
visible data sets.}
\label{flujo_gran_intergran}
\end{figure*}

\subsection{Simultaneous inversion of infrared and visible data sets}

When inverting the four spectral lines simultaneously, more physical
constraints are added, leading to a larger reliability of our analysis. 
Figure \ref{pdf_separat} shows the magnetic field strength distribution (histogram and PDF
representations) inferred from the simultaneous inversion of the two different
spectral ranges (dotted line). The sub-kG field strengths dominate the magnetic
field strength distribution at the internetwork. Even if the information of the
\hbox{$630$ nm} lines has been taken into account in the simultaneous inversion,
the magnetic field strength resembles substantially the one obtained with the
isolated analysis of the $1.56$ $\mu$m lines. 

Figure \ref{pdf_gran_intergran_sim} shows the magnetic field strength
distribution on 
granular and intergranular regions. Granules present a histogram that has a
decreasing exponential shape with intrinsically weak fields. Intergranules,
however, present a maxwellian-type behaviour centered on equipartition fields.
The conclusion is, therefore, the same as that obtained from the inversion
of the infrared pair of lines alone. The right panel of Fig.
\ref{pdf_gran_intergran_sim} shows the distribution of the magnetic filling
factor. Most of the inferred magnetic fields fill up only $2$\% of the
resolution element.

The spatial distribution of the magnetic field strength is shown in the left panel
of Fig. \ref{mapa_b_flujo_sim}. 
The scale of variation of the magnetic field strength is $\sim 2-5''$, showing
patches whose size is similar to that found for the magnetic flux density on the right 
panel of Fig. \ref{mapa_b_flujo_sim}. These scales are larger than the spatial resolution and are not
only due to smearing by the seeing conditions. If we compare the two maps, kG
features are concentrated in few locations corresponding to the highest magnetic
flux densities. On the contrary, this conclusion is not reversible, since there are strong
magnetic flux density concentrations associated with hG field strengths
(for instance, positions [17$''$,26$''$] and [$38''$,$14''$]). The spatial
distribution of magnetic field strengths is really close to the one retrieved
from the infrared inversion. Not only 
the simultaneous and the separate infrared magnetic field strength
distributions are substantially similar, but also 
their spatial distributions are in very good agreement. The spatial distribution
of the magnetic 
flux density is very similar to the one retrieved from the separate inversion.
This is an interesting point since while \hbox{$630$ nm} lines are not capable of
giving good results for the magnetic field strength, they 
are good indicators of the magnetic flux density. 

The left panel of Fig. \ref{theta_camp} shows the magnetic field strength as a function
of the inclination of the magnetic field vector. It seems that there is a tendency for
intrinsically stronger fields to be more vertical. Intrinsically weaker ones
have a wider range of angles. The right panel of Fig.
\ref{theta_camp} represents the histogram of the magnetic flux density. Both
polarities are equally probable in our data set. In fact, the total flux density
in the field of view is \hbox{$-0.055$ Mx/cm$^2$}. This very small flux imbalance has
been considered to point towards a $local~dynamo$ as the mechanism that generates
the magnetic fields on the internetwork \citep{wang_95,lites_02}. The total
unsigned flux in the field of view is $4.10$ G, something 
in between the figure computed from the separate infrared (2.94 Mx/cm$^2$) and
visible (4.85 Mx/cm$^2$) inversions. Figure \ref{flujo_gran_intergran} shows the
magnetic flux density (signed and unsigned) in granular and intergranular
regions. There are no clear differences between the two distributions in
granules or intergranules. It seems that the magnetic flux density remains the
same in the whole solar surface, irrespective of the plasma motions. 


\section{Conclusions}

Simultaneous and co-spatial observations of Fe\,{\sc i} $1.56$ $\mu$m and
Fe\,{\sc i} $630$ nm pair of lines have been analysed. A total of 93\% of the 
observed field of view has polarimetric signals above three times the noise level. 
This means that at 1$''$ spatial resolution the internetwork is full of magnetic fields.

In order to study the internetwork magnetism, the information of the full Stokes
vector has to be taken into account. At this point the $1.56$ $\mu$m lines
present a big advantage when compared to the 630 nm ones. Both pairs of lines are
equally sensitive to the circular polarization. However, the infrared lines
show conspicuous linear polarization signals while the visible ones show these
signals only in few weak features. 

The Stokes V profiles are quite asymmetric at both spectral ranges, having
similar distributions of area and amplitude asymmetries. Most of the observed two-lobed profiles have an amplitude asymmetry of about 15\% and an area
asymmetry mainly zero.

We have performed the separate inversion of the infrared and visible data in order to check their compatibility. Our conclusions are:
\begin{itemize}
\item The magnetic flux density recovered from the independent inversion is the
same at both spectral ranges (both spatial distributions have a strong
correlation). Small differences can be produced by the different formation
heights of the two pairs of lines.
\item There are no systematic opposite polarities for co-spatial Stokes V
profiles in both observations. 
\item The magnetic field strength distributions are very different. The infrared
observations result in equipartition or even weaker fields while the visible
lines show fields in the kG regime.
\end{itemize}

These three points are compatible since the validity of the magnetic
field strength distributions retrieved from the 630 nm data has been questioned.
We conclude that the information of our observations is compatible in both spectral ranges. 
This means that both the $1.56$ $\mu$m and 630 nm lines trace the same magnetic structure 
on the solar surface.
Anyway, it does not mean that the infrared distribution of magnetic field
strengths is the real one. One should keep in mind that the Zeeman effect  is only
capable of detecting magnetic fields in $\approx \sim$2\% of the resolution element. 
What we really know is
that our observations lead to reliable results and that the magnetic field
strengths that we infer are really present on the internetwork. Ram\' irez V\'elez, L\'opez Ariste \& Semel (private comunication),
with observations of a Mn\,{\sc i} line in the visible that is sensitive to the
magnetic field strength, have concluded that the magnetic field strengths on the
internetwork are well below the kG regime. This supports the fact that the infrared is
not biased towards weak fields. 

In order to put more physical constraints, we have performed the simultaneous inversion
of both data sets. The inferred magnetic field strength distribution is
dominated by sub-kG fields. The occupation fraction of those fields 
peaks at $2$\% but with a broad distribution. Two different magnetic field strength 
distributions are present
on granules and intergranules. The granular one has intrinsically weaker field
strengths than the intergranular one, which is centered on the equipartition field
of the photosphere. The magnetic flux density distribution is the same on
granules and intergranular lanes. There is no net magnetic flux density in the
studied field of view, implying that there is a perfect cancellation of positive
and negative magnetic flux density. 

The unsigned magnetic flux density histograms peak at \hbox{$\sim 10$
Mx/cm$^2$}. \cite{martin_87} and \cite{wang_95} obtained the same value using
magnetograms with $2-3''$ spatial resolution. This is agreement with \cite{lites_04},
who show that
there is no apparent increment in the magnetic flux density when improving the spatial
resolution from 1 to 0.6$''$. 

Since the Zeeman effect allows us to study $\sim$ 2\% of
the resolution element, what happens in the rest of the surface? Is
it field-free? If not, which kind of magnetic fields can we find in it?
\cite{andres_07} show the first experimental evidence of magnetic flux
cancellation at the internetwork. This supports the fact that the peak of the
circular and linear polarization histograms well above the noise level is
produced by cancellations in the resolution element. This means that there are 
still undetected magnetic fields at a
spatial resolution of 1$''$ (mixed polarities or extremely weak fields). 
The study of the internetwork magnetism 
has become a very interesting field that deserves more efforts to know if
it can play an important role in the solar global magnetism as claimed by
\cite{javier_04}.

\begin{acknowledgements}
This research has been funded by the Spanish Ministerio de Educaci\'on y Ciencia
through project AYA2004-05792. This article is based on observations taken with
the VTT telescope operated on the island of Tenerife by the Kiepenheuer-Institut
f\"ur Sonnenphysik in the Spanish Observatorio del Teide of the Instituto de
Astrof\'{\i}sica de Canarias (IAC). Many thanks are due to A. Asensio Ramos, I.
Dom\' inguez Cerde\~na, E. V. Khomenko and J. Trujillo Bueno for really
helpful discussions. The authors are specially grateful to A. L\'opez Ariste for
encouraging discussions.
\end{acknowledgements}


\begin{thebibliography}{42}
\expandafter\ifx\csname natexlab\endcsname\relax\def\natexlab#1{#1}\fi

\bibitem[{{Asensio Ramos} {et~al.}(2007{\natexlab{a}}){Asensio Ramos}, {Mart\'
  inez Gonz\'alez}, {L\'opez Ariste}, {Trujillo Bueno}, \&
  {Collados}}]{andres_07}
{Asensio Ramos}, A., {Mart\' inez Gonz\'alez}, M.~J., {L\'opez Ariste}, A.,
  {Trujillo Bueno}, J., \& {Collados}, M. 2007{\natexlab{a}}, ApJ, 659, 829

\bibitem[{{Asensio Ramos} {et~al.}(2007{\natexlab{b}}){Asensio Ramos},
  {Martinez Gonzalez}, \& {Rubino-Martin}}]{andres_marian_07_astroph}
{Asensio Ramos}, A., {Martinez Gonzalez}, M.~J., \& {Rubino-Martin}, J.~A.
  2007{\natexlab{b}}, ArXiv e-prints, 709

\bibitem[{{Asensio Ramos} {et~al.}(2007{\natexlab{c}}){Asensio Ramos},
  {Socas-Navarro}, {L\'opez Ariste}, \& {Mart\' inez
  Gonz\'alez}}]{andres_hector_07}
{Asensio Ramos}, A., {Socas-Navarro}, H., {L\'opez Ariste}, A., \& {Mart\' inez
  Gonz\'alez}, M.~J. 2007{\natexlab{c}}, ApJ, 660, 1690

\bibitem[{{Ballesteros} {et~al.}(1996){Ballesteros}, {Collados}, {Bonet},
  {Lorenzo}, {Viera}, {Reyes}, \& {Rodr�uez Hidalgo}}]{ballesteros_96}
{Ballesteros}, E., {Collados}, M., {Bonet}, J.~A., {et~al.} 1996, A\&A, 115,
  353

\bibitem[{{Beck} {et~al.}(2005{\natexlab{a}}){Beck}, {Schlichenmaier},
  {Collados}, {Bellot Rubio}, \& {Kentischer}}]{beck_schlichenmaier_05}
{Beck}, C., {Schlichenmaier}, R., {Collados}, M., {Bellot Rubio}, L., \&
  {Kentischer}, T. 2005{\natexlab{a}}, A\&A, 443, 1047

\bibitem[{{Beck} {et~al.}(2005{\natexlab{b}}){Beck}, {Schmidt}, {Kentischer},
  \& {Elmore}}]{beck_05}
{Beck}, C., {Schmidt}, W., {Kentischer}, T., \& {Elmore}, D.
  2005{\natexlab{b}}, A\&A, 437, 1159

\bibitem[{{Bellot Rubio} \& {Collados}(2003)}]{luis_03}
{Bellot Rubio}, L.~R. \& {Collados}, M. 2003, A\&A, 406, 357

\bibitem[{{Casini} \& {L\'opez Ariste}(2003)}]{casini03}
{Casini}, R. \& {L\'opez Ariste}, A. 2003, in Solar Polarization 3, ed.
  J.~Trujillo~Bueno \& J.~S\'anchez~Almeida, ASP Conference, 98

\bibitem[{{Collados}(1999)}]{manolo99}
{Collados}, M. 1999, in Third Advances in Solar Physics Euroconference, ed.
  B.~Schmieder, A.~Hofmann, \& J.~Staude, 184 (ASP Conference), 3--22

\bibitem[{{Dom\' inguez Cerde\~na} {et~al.}(2003){Dom\' inguez Cerde\~na},
  {S\'anchez Almeida}, \& {Kneer}}]{ita_jorge_03}
{Dom\' inguez Cerde\~na}, I., {S\'anchez Almeida}, J., \& {Kneer}, F. 2003,
  A\&A, 407, 741

\bibitem[{{Dom\' inguez Cerde\~na} {et~al.}(2006){Dom\' inguez Cerde\~na},
  {S\'anchez Almeida}, \& {Kneer}}]{ita_06}
{Dom\' inguez Cerde\~na}, I., {S\'anchez Almeida}, J., \& {Kneer}, F. 2006,
  ApJ, 646, 1421

\bibitem[{{Howard} \& {Stenflo}(1971)}]{howard_71}
{Howard}, R. \& {Stenflo}, J.~O. 1971, SoPh, 22, 402

\bibitem[{{Keller} {et~al.}(1994){Keller}, {Deubner}, {Egger}, {Fleck}, \&
  {Povel}}]{keller_94}
{Keller}, C.~U., {Deubner}, F., {Egger}, U., {Fleck}, B., \& {Povel}, H. 1994,
  A\&A, 286, 626

\bibitem[{{Khomenko} {et~al.}(2003){Khomenko}, {Collados}, {Solanki}, {Lagg},
  \& {Trujillo Bueno}}]{khomenko_03}
{Khomenko}, E.~V., {Collados}, M., {Solanki}, S.~K., {Lagg}, A., \& {Trujillo
  Bueno}, J. 2003, A\&A, 408, 1115

\bibitem[{{Khomenko} {et~al.}(2005){Khomenko}, {Shelyag}, {Solanki}, \&
  {V\"ogler}}]{khomenko_shelyag_05}
{Khomenko}, E.~V., {Shelyag}, S., {Solanki}, S.~K., \& {V\"ogler}, A. 2005,
  A\&A, 442, 1059

\bibitem[{{Landi degl'Innocenti} \& {Landolfi}(2004)}]{egidio}
{Landi degl'Innocenti}, E. \& {Landolfi}, M. 2004, Polarization in Spectral
  Lines (Kluwer Academic Publishers)

\bibitem[{{Lin}(1995)}]{lin_95}
{Lin}, H. 1995, ApJ, 446, 421

\bibitem[{{Lin} \& {Rimmele}(1999)}]{lin_99}
{Lin}, H. \& {Rimmele}, T. 1999, ApJ, 514, 448

\bibitem[{{Lites}(2002)}]{lites_02}
{Lites}, B.~W. 2002, ApJ, 573, 431

\bibitem[{{Lites} {et~al.}(1993){Lites}, {Elmore}, {Seagraves}, \&
  {Skumanich}}]{lites_93}
{Lites}, B.~W., {Elmore}, D.~F., {Seagraves}, P., \& {Skumanich}, A.~P. 1993,
  ApJ, 418, 928

\bibitem[{{Lites} {et~al.}(1999){Lites}, {Rutten}, \& {Berger}}]{lites_99}
{Lites}, B.~W., {Rutten}, R.~J., \& {Berger}, T.~E. 1999, ApJ, 517, 1013

\bibitem[{{Lites} \& {Socas-Navarro}(2004)}]{lites_04}
{Lites}, B.~W. \& {Socas-Navarro}, H. 2004, ApJ, 613, L600

\bibitem[{{L\'opez Ariste} {et~al.}(2006){L\'opez Ariste}, {Mart\' inez
  Gonz\'alez}, \& {Ram\' irez V\'elez}}]{arturo_07}
{L\'opez Ariste}, A., {Mart\' inez Gonz\'alez}, M.~J., \& {Ram\' irez V\'elez},
  J.~C. 2006, A\&A, 464, 351

\bibitem[{{Mart\' inez Gonz\'alez} {et~al.}(2006){Mart\' inez Gonz\'alez},
  {Collados}, \& {Ruiz Cobo}}]{marian_06}
{Mart\' inez Gonz\'alez}, M.~J., {Collados}, M., \& {Ruiz Cobo}, B. 2006, A\&A,
  456, 1159

\bibitem[{{Martin}(1987)}]{martin_87}
{Martin}, S.~F. 1987, SoPh, 117, 243

\bibitem[{{Muglach} \& {Solanki}(1992)}]{muglach_92}
{Muglach}, K. \& {Solanki}, S.~K. 1992, A\&A, 263, 301

\bibitem[{{Orozco Suarez} {et~al.}(2007){Orozco Suarez}, {Bellot Rubio}, {del
  Toro Iniesta}, {Tsuneta}, {Lites}, {Ichimoto}, {Katsukawa}, {Nagata},
  {Shimizu}, {Shine}, {Suematsu}, {Tarbell}, \& {Title}}]{david_07_astroph}
{Orozco Suarez}, D., {Bellot Rubio}, L.~R., {del Toro Iniesta}, J.~C., {et~al.}
  2007, ArXiv e-prints, 710

\bibitem[{{Reardon}(2006)}]{reardon_06}
{Reardon}, K.~P. 2006, SoPh, 239, 503

\bibitem[{{Rees} \& {Ying}(2003)}]{rees03}
{Rees}, D. \& {Ying}, G. 2003, in Solar Polarization 3, ed. J.~Trujillo~Bueno
  \& J.~S\'anchez~Almeida, 307 (ASP Conference), 85--97

\bibitem[{{Ruiz Cobo} \& {del Toro Iniesta}(1992)}]{basilio_92}
{Ruiz Cobo}, B. \& {del Toro Iniesta}, J.~C. 1992, ApJ, 398, 375

\bibitem[{{S\'anchez Almeida} {et~al.}(2003){S\'anchez Almeida}, {Dom\' inguez
  Cerde\~na}, \& {Kneer}}]{jorge_ita_03}
{S\'anchez Almeida}, J., {Dom\' inguez Cerde\~na}, I., \& {Kneer}, F. 2003,
  ApJ, 597, L177

\bibitem[{{S\'anchez Almeida} \& {Landi Degl'Innocenti}(1996)}]{jorge_96}
{S\'anchez Almeida}, J. \& {Landi Degl'Innocenti}, E. 1996, SoPh, 164, 203

\bibitem[{{S\'anchez Almeida} \& {Lites}(2000)}]{jorge_00}
{S\'anchez Almeida}, J. \& {Lites}, B.~W. 2000, ApJ, 532, 1215

\bibitem[{{Schlichenmaier} \& {Collados}(2002)}]{schlichenmaier_02}
{Schlichenmaier}, R. \& {Collados}, M. 2002, A\&A, 381, 668

\bibitem[{{Skumanich} \& {L\'opez Ariste}(2002)}]{skumanich_02}
{Skumanich}, A. \& {L\'opez Ariste}, A. 2002, ApJ, 570, 379

\bibitem[{{Socas-Navarro}(2003)}]{hector03}
{Socas-Navarro}, H. 2003, in Solar Polarization 3, ed. J.~Trujillo~Bueno \&
  J.~S\'anchez~Almeida, ASP Conference, 330--335

\bibitem[{{Socas-Navarro} \& {Lites}(2004)}]{hector_04}
{Socas-Navarro}, H. \& {Lites}, B.~W. 2004, ApJ, 616, 587

\bibitem[{{Socas-Navarro} \& {S\'anchez Almeida}(2002)}]{hector_02}
{Socas-Navarro}, H. \& {S\'anchez Almeida}, J. 2002, ApJ, 565, 1323

\bibitem[{{Solanki} \& {Stenflo}(1984)}]{sami_84}
{Solanki}, S.~K. \& {Stenflo}, J.~O. 1984, A\&A, 140, 185

\bibitem[{{Stenflo}(1973)}]{stenflo_73}
{Stenflo}, J.~O. 1973, SoPh, 32, 41

\bibitem[{{Trujillo Bueno} {et~al.}(2004){Trujillo Bueno}, {Shchukina}, \&
  {Asensio Ramos}}]{javier_04}
{Trujillo Bueno}, J., {Shchukina}, N., \& {Asensio Ramos}, A. 2004, Nature,
  430, 326

\bibitem[{{Wang} {et~al.}(1995){Wang}, {Wang}, {Tang}, {Lee}, \&
  {Zirin}}]{wang_95}
{Wang}, J., {Wang}, H., {Tang}, F., {Lee}, J.~W., \& {Zirin}, H. 1995, SoPh,
  160, 277

\end{thebibliography}

\end{document}